A tale of two test setups:

The effect of randomizing a popular conceptual survey in electricity and magnetism

Emanuela Ene[1], Department of Physics& Astronomy, Texas A&M University, College Station, Texas 77843-4242, USA

Robert C. Webb[2], Department of Physics & Astronomy, Texas A&M University, College Station, Texas 77843-4242, USA

## *Abstract*

We describe a retrospective study of the responses to the Brief Electricity and Magnetism Assessment (BEMA) collected from a large population of 3480 students at a large public university. Two different online testing setups were employed for administering the BEMA. Our tale focuses on the interpretation of the data collected from these two testing setups. Our study brings new insight and awareness to the effect of the randomization on the BEMA test results.

Starting from an a priori common sense model, we show simple methods to detect and separate guessing from the genuinely thought responses. In addition, we show that the group of responses with low scores (7 or less out of 30) had different answer choice strategies than the groups with average or high scores, and that the time-in-testing is an essential parameter to be reported. Our results suggest that the data should be *cleaned* to insure that only *valid* times-in-testing are included before reporting and comparing statistics. We analyze in detail the effect of the answer-options randomization and the effect of the shuffling of the questions' order. Our study did not detect any significant effect of the option randomization alone on the BEMA scores, for *valid* times-in-testing. We found clear evidence that shuffling the order of the independent questions does not affect the scores of the BEMA test, while shuffling the order of the dependent questions does.

### A. Introduction

Our team has attempted to measure learning gains in the introductory electricity and magnetism course PHYS208 at Texas A&M University (TAMU) using the popular Brief Electricity and Magnetism Assessment (BEMA) [1]. The multiple-choice BEMA has been employed widely in colleges and universities across the USA in either paper form or online [2-5] to evaluate student knowledge levels in introductory courses in electricity and magnetism (E&M). The BEMA questions have from three to ten answer-options per question, out of which only one is correct (the key). The incorrect options generally reflect the common misconceptions of the beginner students in an E&M course.

We have administered the BEMA at TAMU as a non-credit, on-line assignment in the proctored environment of the physics lab during the first and last weeks of the semester. Between Fall 2014 and Summer 2016, for six semesters, we have administered BEMA on the WebAssign platform [6]. Between Fall 2016 and Summer 2017, for three semesters, we have administered the BEMA questions on a local platform at TAMU [7]. The BEMA-questions were presented differently on these two online platforms, and also differently from the author's paper version of the test [1].

Our main interest at TAMU was to calibrate an objective item bank that could be employed to assess teaching effectiveness in our introductory calculus based E&M course PHYS208, meanwhile building on the existing research evidence pertaining to the BEMA. Between 2013 and 2017, the PHYS208 course at TAMU was taught following two traditional curricula: Don't Panic[8] and University Physics[9]. The content, sequence, and format of the lectures varied across course curricula and lecture instructors. All the students enrolled had to attend weekly three hours of lecture, in a lecture hall accommodating from 100 to 150 persons that corresponded to 3 to 5 sections, and a three-hours recitation/lab session. All the PHYS208 class exams were comprehensive and focused on problem solving. The midterm exams for PHYS208 were common and simultaneous for all the students following the same curriculum (either Don't Panic or University Physics). The lecture instructors wrote and scheduled independently their own final exam. We knew that the BEMA itself was limited in its topical coverage compared to the materials presented in our E&M classes at TAMU.

Our attempt to model these data collected on WebAssign for ordering the BEMA questions as an invariant one-dimensional item[3] scale for knowledge measurement[10] was not successful. Similar to the results obtained at other universities, we found that some BEMA questions are addressing knowledge beyond the electricity and magnetism and thus introduce confounding factors into this physics concept survey[11]. One of these outlier questions from the BEMA was detected by the authors themselves: "Question 9 […] clearly stands out as different from all other questions. This question asks students to select an algebraic expression […] Almost no students get this question correct […]"[1]. We found that nine BEMA questions on WebAssign were answered correctly by less than 20% of our students, and that were many strongly associated pairs of items, more than L. Ding detected when administering the paper version of the test to a 190-student sample[11].

In order to keep track of the answer-option chosen by our students, and to be able to upload the BEMA data on PhysPort [3] where we could make comparisons with the results obtained by our colleagues from other universities, we chose to move the testing off from WebAssign starting with Fall2016. We presented twenty-two BEMA questions in randomized order as independent items on a secure server at TAMU (local platform) during the second stage of our study that spanned one academic year. We did not present the items identified as outliers in our data from WebAssign: Q4, Q5, Q9, Q19, and the groups of strongly associated questions Q14, Q15, Q16, and Q28, Q29. We retained only one group of questions written to be strongly dependent on each other, the three connected questions on Coulomb's law. Because this trio was not presented in sequence, we attached copies of the original diagram to each question of the trio. The positions of these three items within the shuffled BEMA, on the local platform, could be for example 8 for Q3, 16 for Q2, and 26 for Q1.

The WebAssign testing platform was recommended by the BEMA's authors, who employed it for the statistical evaluation of the BEMA in 2003 [1]. At the time of our study at TAMU, the questions on WebAssign were listed in fixed order all on a single web page. The test-takers could scroll through and select an answer-option by either clicking a radio button followed by a verbal or symbolic statement, for 13 questions, or by selecting from an alphabet letter (i.e., A, B, C, etc.) from a dropdown list for the other 18 questions. In addition, the WebAssign team randomized the order of the answer-options for nine BEMA questions. For example, five of the six answer-options of the question Q7 appeared at a different position in the list (at A, B, C, etc.) each time the test was accessed on WebAssign. Figure 1, which is an excerpt from the gradebook of the summer 2016 semester containing twelve answers to the BEMA question Q7, illustrates the effect of the options' randomization on the gradebooks. In Figure 1, the alphabet letters from A to F in the column "response" represent the answer-option chosen by students. For each possible option choice, the column labeled "correct" lists a 1 (correct) only if the correct answer (the key) happens to be at that position during randomization. The chance that the answer-option A from the paper version would land on the position A of the randomized list is about 1/5.

---

[3] We call a question of the test "item" for emphasizing its statistical properties.

In Figure 1, the column labeled "section" represents the numerical code of a group of 20 to 30 students who attended the same recitation/lab section. The observable "section" linked the students' responses to their instructors; the instructors received the anonymous test statistics to help them tailor their teaching to the needs of their sections of students.  The column labeled "score" represents here the total score for the 31 questions in BEMA. The column labeled "TAQ_minutes" represents the total time-in-testing[4], an observable that proved to be essential for explaining the scores and the responses (option choices).

On the local platform at TAMU, the BEMA questions were presented in randomized order, one question per page, while the order of the answer-options for each question was kept fixed. The test-takers selected an answer-option by clicking a radio button followed by a verbal or symbolic statement, consistently for all the BEMA questions. We shuffled the original sequencing of the BEMA questions in order to transform them into independent questions.

### B.  Motivation

As the BEMA has been validated [1, 12] for its questions presented in fixed order from 1 to 31, and for a fixed order of the answer-options, we asked ourselves if the scores are affected by presenting the answer-options in random order, such as it was the case on the WebAssign platform. We believed that the score differences accompanying the change in questions' presentation could come from those students who randomly guess the answers but not from those students who think through their answers[5][13]. In addition, we asked ourselves if the responses[6] obtained from the BEMA testing under different test presentation setups, i.e. all questions on one page vs. one question per page with no back navigation, can be equated to each other.

We were aware of a plethora of studies and foundational books discussing the possible effects of rearranging the answer options, of the arrangement of the conceptual questions in a different order than they were presented in the text book, and discussing statistical methods for detecting significant differences in the data collected during online testing [14-17]. We did not find any previous study on questions or options randomization for the BEMA therefore we believe that our study would bring new insight and awareness to the effect of the randomization on the BEMA test results.

### C.  Data collection

In this study, all testing took place online in the proctored environment of a computer lab, with the completion time restricted to 60 minutes[7]. No incentive was offered to the students, however it was explained that the goal of the online testing was to objectively measure their knowledge level in order that their instructors could adapt their teaching approach to the specific needs of the students based on the anonymous aggregated test statistics. In addition, no feedback was offered to students after the testing. The pre-instruction test was offered during the first week of the semester and the post-instruction test occurred generally one week before the final exam of the semester. Out of the 3480 students that were in this study at TAMU, 63% repeated the pre-post instruction testing during the same semester, 178 (5%) took the BEMA test more than twice because they were enrolled in the PHYS208 course multiple times, and the rest took the test only once either at pre- or at post-instruction.

---

[4] Time-in-testing represents how many seconds (or minutes) have students spent since they accessed until they submitted the online test. The abbreviation on WebAssign is "TAQ", Time to Answer the Question.

[5] Prior studies have shown that people mark preferentially the answer options A, B, or C in questionnaires with  blank questions (ref. 13.  van Heerden, J. and J. Hoogstraten, *Response Tendency in a Questionnaire without Questions.* Applied Psychological Measurement, 1979. **3**(1): p. 117-121. )

[6] Throughout this paper, we call "response" the position in the options' list chosen and submitted by a person, disregarding the actual content of the answer at that position in the options' list. We will also call "response" the collection of option choices for all the questions in the test submitted by a person at one instance.

[7] The generous interval of 60min accommodated our disability students, and allowed time for troubleshooting if computers malfunctioned.

The BEMA questions presented in fixed order on WebAssign were graded as prescribe by the authors[8]. The answers to the shuffled questions on the local platform were consistently graded 1 for correct, 0 for incorrect, and NA if no answer submitted.

| student | course | section | assignment_ | score | TAQ_minutes | question | response | correct |
|---|---|---|---|---|---|---|---|---|
| 1 | Phys 208 | 301 | Pre-course - | 6 | 17 | 7 | A | 1 |
| 2 | Phys 208 | 301 | Post-course | 9 | 14 | 7 | A | 0 |
| 3 | Phys 208 | 307 | Pre-course - | 9 | 10 | 7 | B | 0 |
| 4 | Phys 208 | 307 | Pre-course - | 11 | 15 | 7 | B | 1 |
| 5 | Phys 208 | 301 | Pre-course - | 6 | 3 | 7 | C | 0 |
| 6 | Phys 208 | 302 | Pre-course - | 15 | 26 | 7 | C | 1 |
| 7 | Phys 208 | 307 | Post-course | 17 | 11 | 7 | D | 1 |
| 8 | Phys 208 | 307 | Pre-course - | 5 | 17 | 7 | D | 0 |
| 9 | Phys 208 | 308 | Pre-course - | 8 | 4 | 7 | E | 0 |
| 10 | Phys 208 | 308 | Pre-course - | 8 | 11 | 7 | E | 1 |
| 11 | Phys 208 | 308 | Post-course | 3 | 3 | 7 | F | 0 |
| 12 | Phys 208 | 308 | Pre-course - | 2 | 2 | 7 | F | 0 |

**Figure 1 – Example of answer-option randomization of the BEMA on WebAssign**
This excerpt from the gradebook for BEMA testing during the summer semester of 2016 shows the answers to the question Q7 submitted by twelve different students at either pre- or post-course testing. Column "score" represents the total score for the 31 questions in BEMA, "TAQ_minutes" represents the total time-in-testing, column "response" gives the option chosen by the students from that (randomized) sequence presented to them, column "correct" shows 0 for incorrect and 1 for correct answers.

### D. Data grouping

Before starting the data analysis, we constructed an a priori model based on three common sense statements rooted in folklore and our teaching experience:

i. Students who do not read or do not understand the question but submit an answer have the tendency to randomly choose one of the first options in the list, independently of the number of options presented for that question.

ii. As a consequence of the previous hypothesis, the frequency of choosing the options A, B, and C should stay the same for the questions that are equally unknown to students, no matter the presentation setup (i.e., all questions on one page or each question on its page).

iii. The responses corresponding to short test taking durations are most likely guessed.

---

[8] The BEMA contains three groups of dependent questions. At the author's prescription, the answer for the questions Q3 and Q16 receives a score of 1 only if it is consistent with the answers for the preceding questions in that group. Only if both answers for the questions Q28 and Q29 are correct, their cumulative score is 1; the cumulative score for Q28 and Q29 is 0 if at least one answer is incorrect. The total maximum score for the 31 questions of BEMA is 30. The answers for Q3 and Q29 were incorrectly scored by WebAssign between 2014 and 2017, receiving 1 point if correct regardless the answers for Q2 and Q28. This grading issue affected all the users of WebAssign, not only our TAMU users. We have regraded all the BEMA tests taken by our students on WebAssign for this analysis

A tale of two test setups: The effect of randomizing a popular conceptual survey in electricity and magnetism

The present study analyzes the answer choices for eighteen BEMA questions. Nine of these questions had the answer-options randomized on WebAssign, but in fixed order on the local server. The other nine questions had the answer-options in fixed order for both test presentations, and were selected as a control group for the effect of option randomization. Section D, subsection c) will clarify our rationale for selecting the questions in the control group.

We introduced categorical variables (data groupings) that allowed a quantitative analysis of the statements of our a priori model. The categorical variables, which trends we analyzed in our study, were: type of diagnostic, test take, score quartiles, question-group, difficulty, and time-category. We will describe in detail the characteristics of these categories in the sections that follow.

a) Categories based on the calendar date when the test was taken – Diagnostic and Take

From the point of view of the calendar date, we assigned the data to one of two types of knowledge diagnostic, pre-instruction or post-instruction. The data were also linked to the individual persons sitting for testing only once (PREonly or POSTonly), twice (repeat) testing during the same school semester, or more than twice during multiple semesters. Table 1-1 gives the person distribution across the category "Take".

**Table 1-1 Categories of persons participating in the BEMA testing**

| Test-takes | Take-category | | | | Sums | Proportion (%) |
|---|---|---|---|---|---|---|
| | Multiple | POSTonly | Repeat | PREonly | | |
| 1 | 0 | 68 | 0 | 1046 | 1114 | 32.0 |
| 2 | 0 | 0 | 2188 | 0 | 2188 | 62.9 |
| 3 | 131 | 0 | 0 | 0 | 131 | 3.8 |
| 4 | 41 | 0 | 0 | 0 | 41 | 1.2 |
| 5 | 6 | 0 | 0 | 0 | 6 | 0.2 |
| Sums | 178 | 68 | 2188 | 1046 | 3480 | 100.0 |

b) Categories based on person's score – Score Quartiles

We assumed that the total score, for the questions presented at one instance, is a sufficient statistic to characterize a persons' knowledge at that moment. Although the distribution of the total scores was semester dependent, the score distributions for each of the three academic years of this study were strikingly alike.

The mean scores of the 4108 submitted sets of responses on WebAssign during the first six academic semesters of our study were 27%±10% at pre-instruction and 42%±18% at post-instruction, very similar to those reported by other universities [1, 4, 5]. The overall mean score for the BEMA test on WebAssign, when the pre- and post-instruction data were pooled together, was 10.62 (35.4%).

We assigned the submitted responses to four bins (quartiles) based on the overall raw score[9] distribution. We chose the cutoff scores of 7, 10, and 14 to obtain a roughly equal number of responses submitted in each bin (quartile). The cutoffs represented the score percentages 23.33%, 33.33%, and 43.33% out of the maximum total score of 30. In the lower quartiles, 73% of the scores came from pre-instruction and 27% from post-instruction testing. In the upper quartiles, 42% of the scores came from the pre- and 58% from the post-instruction testing. A significant number of students scoring in the lower quartiles at pre-instruction repeated the test at the end of the semester, showed improved knowledge, and scored in the upper quartiles at post-instruction.

---

[9] The "raw scores" were automatically stored by the online gradebook. They came from all the categories of test-takers.

c)   Categories based on the position of the correct answer in the sequence of options – Question Group

We grouped the BEMA questions into four categories (Question Groups) based on the position of the correct answer (key), the number of answer-options, and the option randomization on WebAssign. **Table 1-2** gives these question groups. The nine questions with the options in randomized order on WebAssign are grouped on the left-hand side of **Table 1-2**. "**Group 1**" contains three questions with three to five answer-options and the fixed key at *A*; on WebAssign the floating correct answer could be *A,B, C, or D*. "**Group 2**" contains six questions with five to nine options; the fixed key is at *D, E, F,* or *G* but it floated from A to I on WebAssign.

The nine questions with the options in fixed order on WebAssign are grouped on the right-hand side of **Table 1-2** . "**Group 3**" contains questions with seven to ten answer-options; four of them have the fixed key at A and one of them has the fixed key at B. This is the control for **"Group 1".**   **"Group 4"** contains four questions with six to eight options and with the key fixed at the same position, somewhere in the second half of the option list, at *D, E, F,* or *G*. This is the control for **"Group 2".**  Note that all questions presented on the local platform were in a shuffled order but their answer choices match the descriptions of groups 1 through 4 above.

**Table 1-2   The eighteen BEMA questions analyzed in this study.** The questions were listed: †= in the original order, all questions on the same page; ‡= in shuffled order, one question per page. When the options were listed in fixed order, the key was positioned at: i= A or B; ii= D, E, F, or G. *= this question was offered only on WebAssign.

| Options listed in randomized order on WebAssign† but in fixed order on the local platform‡ | | | | | Options listed in fixed order both on WebAssign† and the local platform ‡ | | | |
|---|---|---|---|---|---|---|---|---|
| Group | Question name | Question difficulty | Number of options | | Group | Question name | Question difficulty | Number of options |
| | | | total | randomized | | | | |
| $1^i$ | | | | | $3^i$ | Q1 | easy | 7 |
| | Q8 | easy | 3 | 2 | | Q2 | easy | 7 |
| | Q18 | easy | 4 | 3 | | Q3 | easy | 9 |
| | Q9* | hard | 5 | 5 | | Q21 | medium | 10 |
| | | | | | | Q24 | hard | 8 |
| $2^{ii}$ | Q13 | medium | 5 | 4 | $4^{ii}$ | | | |
| | Q7 | hard | 6 | 5 | | | | |
| | Q12 | hard | 7 | 7 | | Q6 | hard | 7 |
| | Q27 | hard | 8 | 7 | | Q25 | hard | 8 |
| | Q10 | hard | 9 | 8 | | Q26 | hard | 7 |
| | Q11 | hard | 9 | 8 | | Q31 | hard | 6 |

d)   Categories based on item popularity (proportion of correct answers for that question) – Question Difficulty

We calculated the item popularity[10] as the proportion of correct answers out of the total submitted answers to one question. The item popularities calculated over the pooled bin of responses submitted by the students in this study are given in [18] Appendix 1. We labeled a question as "easy" if its popularity calculated solely for quartile 2 was greater than 50% (more than half of the answers correct), "medium" if its popularity was under 50% for quartile 2 but greater than 50% for the quartiles 3 and 4, or "hard" if its popularity for quartile 3 was under 50% (more than half of the answers incorrect). The reader should keep in mind that the parameters "item popularity" and "difficulty" reflect the population of students and the method of instruction at TAMU, and the testing method employed during this study.

---

[10] Item popularity is a sample dependent statistic that gives a rough estimate of the proficiency of the average person with the concepts and skills pertaining to a particular question in the test.

   e) Categories based on the time-in-testing – Time Categories

The average time in testing for 31 questions was 17.1 min on WebAssign. Data analysis showed that the rate of submitting answers decreased abruptly for times-in-testing greater than 7 minutes; the 7 min duration corresponds to an average of 13.5 seconds to read and submit an answer. We identified a significant linear dependence between the total score and the time-in-testing for those persons spending on average less than 13.5 seconds per question.

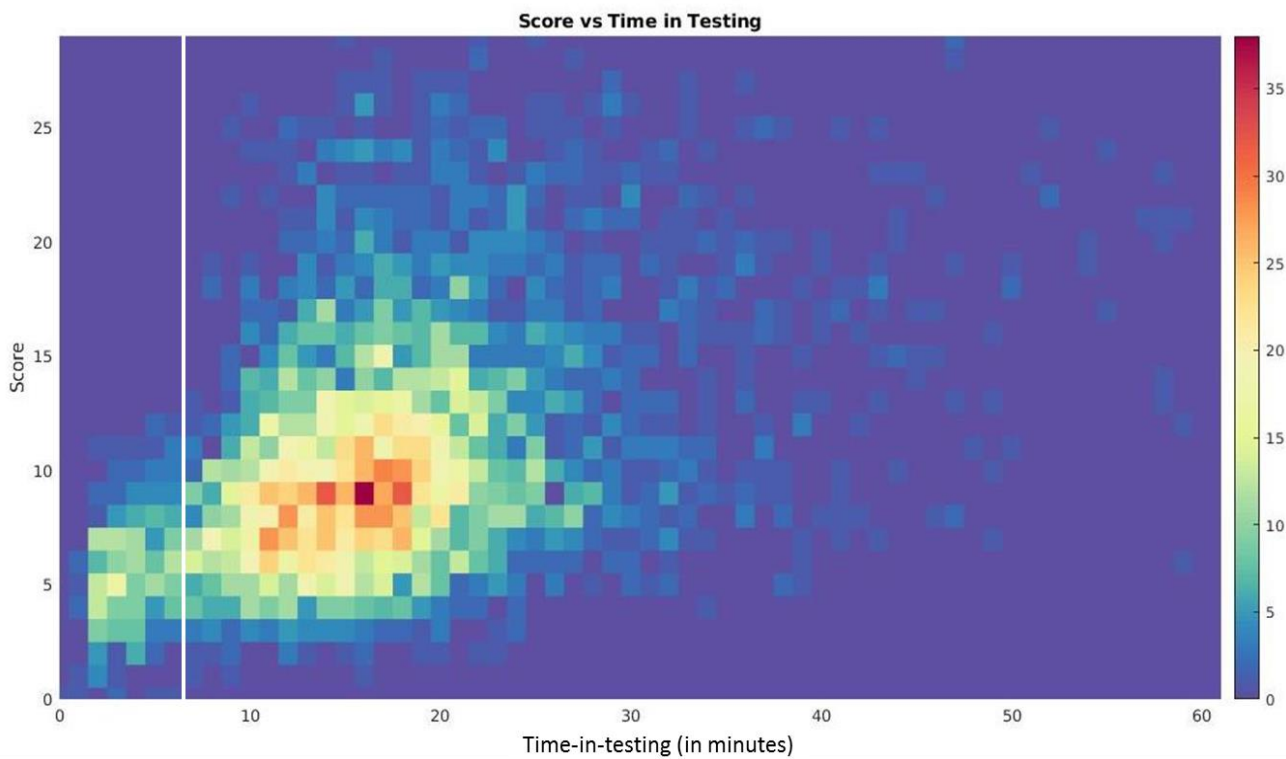

**Figure 2 – BEMA scores vs. time-in-testing, for data collected on WebAssign.**
The testing was administered online on WebAssign between Fall14 and Summer16. A vertical line at t=7 min represent an approximate boundary between two groups of submissions, one centered on scores of 5 and time-in-testing of 4 min, the other on scores of 9 and time-in-testing of 16 min. The data points for short times account for 323 persons, the data points for longer times account for 3786 persons. We inferred that the persons in the two groups had employed different strategies to answer the BEMA questions. We labeled the group of data to the left of the time-boundary *quick* and the group to the right *valid*.

Figure 2 illustrates the two different answering trends separated by a boundary at 7 minutes. In the 3D plot in Figure 2, because the time-in-testing was rounded to the nearest integer, multiple persons share the same (time, score) point of the plot. The multiplicity is indicated by the color. The reader can see two centers of data density, one around the score of 5 and time of 4 min, the other around the score of 9 and time of 16 min. The entries in the group with short times account for 323 persons, while the entries for longer times account for 3786 persons.

For short times-in-testing, the rate answering was of one correct answer submitted each two minutes. The ratio (score/time) had a large variability for durations greater than 7 minutes, as one would expect from a diverse population; the average trend was one correct answer each six minutes for this group. We inferred that the persons in these two groups, to the left and to the right of the 7 min boundary, had employed different strategies in answering test questions.

We found a similar time-categories dependence of the responses in the case of the test presented on the local TAMU platform, with one question per page. As a result of this different test format, it took a longer time for students to navigate through questions on the local platform than on WebAssign. The average time-in-testing was 22.3 min and the time boundary for the linear dependence score vs. time was at 8.8 min. For times-in-testing shorter than 8.8 min, we found an average rate of one correct answer each 1.6 min, faster than when all questions were on same page. The faster rate likely reflected the fact that the *quicks* did not navigate through all the pages on the local platform and submitted answers to only a few questions before exiting the assessment. For one question per page presentation, we found a large variability of the ratio (score/time) for times-in-testing greater than the time-boundary of 8.8 min. The average trend, for valid times, was of one correct answer in seven minutes. Thus, whether the test setup required test takers to scroll down through a single page or to navigate through numerous pages, some persons managed to answer the questions at a fast pace.

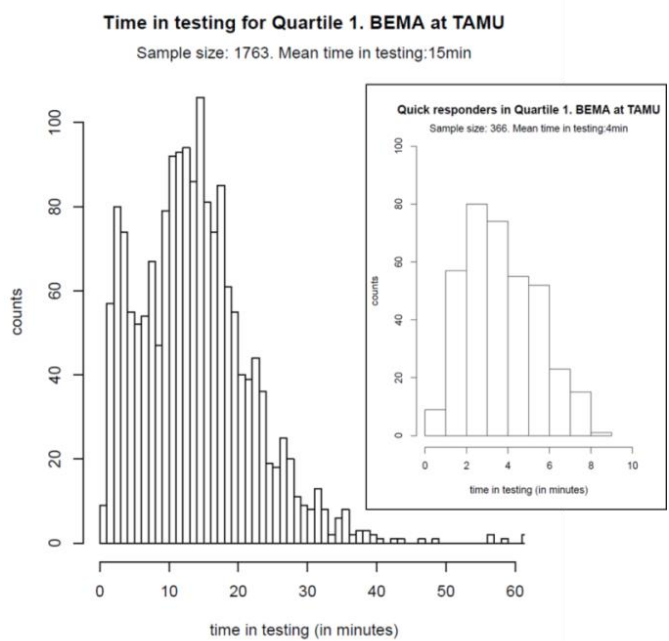

**Figure 3 – Distribution of the time-in-testing for responses scored in quartile 1.** The distribution is bimodal, with a peak at 4 min (time-category *quick*) and another at 15 min (time-category *valid*). The inset shows the time distribution of the 366 *quick* responders from both WebAssign and local platform.

We grouped the responses corresponding to short times in the time-category *quick* and the rest in the time-category *valid*[11]. *Quick* responses accounted for 21% of the submissions in quartile 1, 5% in quartile 2, and 1.2% in quartile 3. No *quick* responses belonged to quartile 4 (scores greater than 14.) Figure 3 shows the distribution of the time-in-testing for responses scored in quartile 1 (low scores). The distribution is bimodal, with a peak at 4min (time-category *quick*) and another at 15 min (time-category *valid*). The inset represents the time distribution of the 366 *quick* responders from both the WebAssign and the local platform.

---

[11] The word "*valid*" relates to the goal of the present study. *Valid* means data coming from a time-in-testing wise homogeneous population and thus suitable for analyzing the effect of the test presentation.

### E. Data Analysis

In checking for statistical significance in our data we will use two methods: the normal approximation method and the chi-squared method. When testing the difference of popularity for significance, we assumed that the item popularity is the best estimator of the probability to answer correctly and applied the normal approximation for the confidence intervals [19]. We considered the popularities equal (no significant difference) if the confidence interval of the difference in popularity spread across the zero value. For each of the eighteen BEMA-questions in this study, we calculated the difference of the popularities obtained from two disjoint samples of responses for a single question, i.e., responses if the answer-options were in fixed order and responses if the answer-options were in randomized order, and these differences were analyzed using the normal approximation method.

Next we calculated the option choice distribution [20] separately for each disjoint sample of responses of interest in our study, i.e., *quick* responses and *valid* responses. We checked if the distributions were significantly different via the chi-square test. To assist the reader, an example of applying the chi-square test for comparing two option choice distributions is presented in [18] Appendix 7.

Table 1-3 summarizes the collection of disjoint groups of responses that we analyzed. We considered 14 disjoint groups containing answers to 18 BEMA questions collected on WebAssign, where the questions were listed in fixed order; another 14 disjoint groups containing answers to 17 BEMA questions were collected on the local platform, where the questions were listed in shuffled order.

### PART I. The effect of short times-in-testing on option choice and score

We began by analyzing the responses corresponding to short test durations. This response category was mostly associated with low scores. For example, in the responses collected on WebAssign, quartile 2 had 53 of *quick* out of 1171 submissions, and quartile 3 had 12 of *quick* out of 904 submissions. The complete distribution of the *quick*s across quartiles and testing platforms is given in [18] Appendix 2. Because the attempt to detect the effects of the time-in-testing in the upper quartiles was overshadowed by the size effects, we focused the analysis of the *quick* responses only in quartile 1.

#### a) Answering patterns rather than random guessing for short times-in-testing

Our data showed that selecting all answers under the same alphabet letter, an approach known as blind guessing, happened for the time category *quick* but not for time category *valid*. We found responses with all A-s submitted in less than 7 min but nobody submitted an all C-s response, as we expected a priori [21]. Instead, we found *quick* responses containing mostly C-s, sprinkled with a couple of A-s or D-s. Other *quick* responders apparently played the game of answering in patterns. This was the case of a student coded Kim (not a real name) who took the BEMA four times during two subsequent academic semesters. At pre-instruction in Spring16, Kim needed more than 7 min for getting a score of 5; the options he chose were mostly A, B, C or the last option in the sequence. At the post-instruction testing of the same Spring16 semester, Kim chose A (the first option in the sequence) for all questions, spent 4 min in testing and made a score of 8. At the third BEMA testing, pre-instruction in Summer2016, Kim chose only the last option in the sequence for all questions and made a score of 2 in 2 min. At the fourth test, post-instruction in Summer2016, Kim chose mostly D-s and some of "none of the above", and made a score of 6 in 2 min. The complete list of Kim's answers is shown in [18] Appendix 3.

A tale of two test setups: The effect of randomizing a popular conceptual survey in electricity and magnetism

*Last updated: 8/9/2019*

**Table 1-3 –Disjoint groups of responses analyzed.** We considered 14 disjoint groups of responses for each testing setup.

| Type of diagnostic | Score Quartile | Time category | Question list | Response group |
|---|---|---|---|---|
| Pre-instruction | 1 | valid | fixed order | 1 |
| | | | shuffled | 2 |
| | | quick | fixed order | 3 |
| | | | shuffled | 4 |
| | 2 | valid | fixed order | 5 |
| | | | shuffled | 6 |
| | | quick* | fixed order | 7 |
| | | | shuffled | 8 |
| | 3 | valid | fixed order | 9 |
| | | | shuffled | 10 |
| | | quick* | fixed order | 11 |
| | | | shuffled | 12 |
| | 4 | valid | fixed order | 13 |
| | | | shuffled | 14 |
| Post-instruction | 1 | valid | fixed order | 15 |
| | | | shuffled | 16 |
| | | quick | fixed order | 17 |
| | | | shuffled | 18 |
| | 2 | valid | fixed order | 19 |
| | | | shuffled | 20 |
| | | quick* | fixed order | 21 |
| | | | shuffled | 22 |
| | 3 | valid | fixed order | 23 |
| | | | shuffled | 24 |
| | | quick* | fixed order | 25 |
| | | | shuffled | 26 |
| | 4 | valid | fixed order | 27 |
| | | | shuffled | 28 |

*=the proportion of the quick responses is negligible for this score quartile

The story of Kim's answers and scores emphasizes that some of the guessers, if not most of them, applied answering patterns and did not pick an answer from the list at random. Generally, our undergraduate students are well trained for guessing strategies; the internet educates them with plenty of examples of maximizing the score in multiple choice tests via blind guessing [22]. As counterintuitive as it may be, the relatively successful score of 8 of the blind guessing in Kim's case was due to the option randomization on WebAssign. The randomization moved the correct option at A for some questions that were written with a fixed key at B, D, E, or F.  Other students who adopted blind guessing "all A-s" were not that lucky and their total score was usually 5.

b) Comparing item popularities for *quick* and *valid* time categories, for questions in fixed order

The answering strategy discussed in the previous section characterized mostly the submissions from quartile 1 (low scores). For the data collected on WebAssign, where the questions were in fixed order, we found significant differences at 0.05-level between the *quick* and the *valid* group of responses for twelve of the eighteen BEMA-questions in this study. The interested reader can find the results of the popularity calculations and the confidence intervals in [18] .

Figure 4 shows the results in graphical form.  The coordinates of a point on the plot in Figure 4 give the item-popularities of the same question for the category *valid* (on the abscissa) and for *quick* (on the ordinate). The shape of the point plotted indicates the group of questions (item number). Those items on or close to the line of equal popularity have similar chances to be answered correctly either *quickly* or in a longer time. Differences under 5% are not significant.

We calculated with 95% confidence that twelve BEMA questions on WebAssign were answered with different probabilities of success by the two time-categories of quartile 1.  Five of these questions had the options in randomized order: Q10, Q12, Q13, Q18, and Q27. Seven of these questions had the options in fixed order: Q1, Q3 graded as a dependent question, Q21, Q6, Q25, Q26, and Q31.

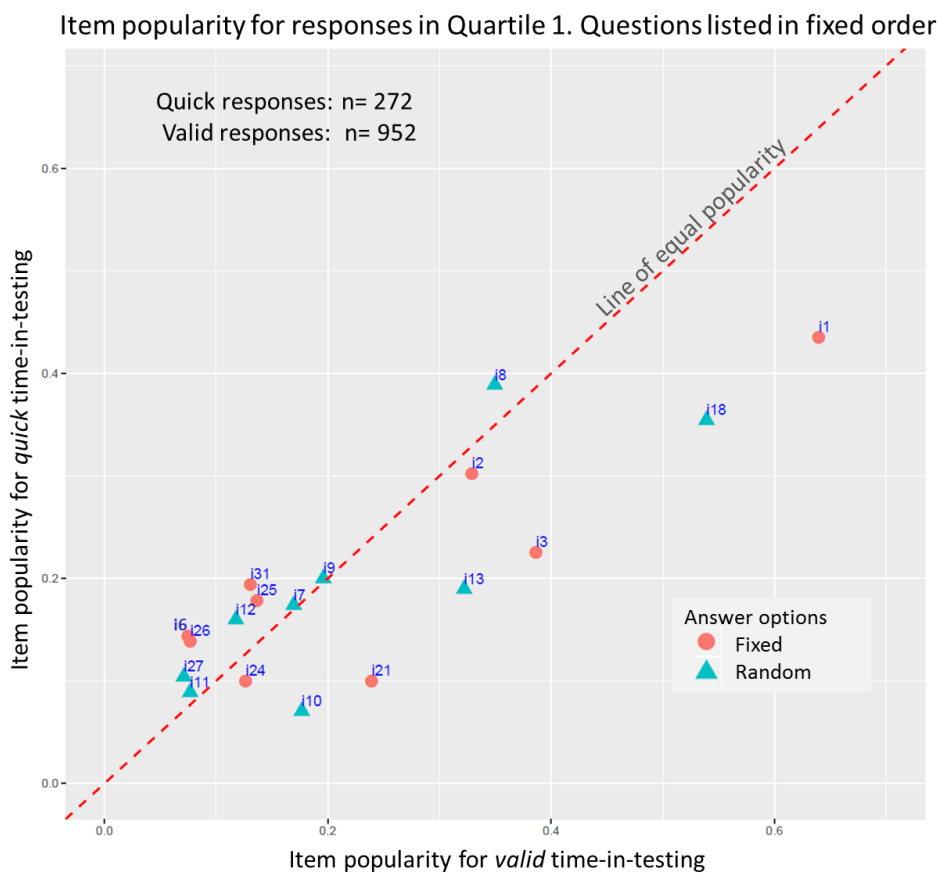

Item popularity for responses in Quartile 1. Questions listed in fixed order

**Figure 4 –Visual comparison between the time-categories *quick* and *valid* in quartile 1 for data from WebAssign, pre- and post-instruction pooled together.** The coordinates of a point of this plot represent the item-popularities of the same question calculated for the categories *valid* (on the abscissa) and *quick* (on the ordinate). Those items on or close to the line of equal popularity have similar chances to be answered correctly either *quick*ly or in a longer time. The shape indicates the question group and the item difficulty is color coded. The *quick*s chose significantly more incorrect answers than the *valid*s on WebAssign for five questions with the options in randomized order (Q10, Q12, Q13, Q18, and Q27) and for seven questions with the options in fixed order (Q1, Q3, Q6, Q21, Q25, Q26, and Q31).

Knowing that a score in quartile 1 can be achieved by blind guessing, and that quartile 1 amasses responses that are mostly incorrect, we speculated for possible causes of the item popularity differences in Figure 4.  Three of the questions, Q1, Q18, and Q3 are easy but have tricky answer-options therefore the students who spent time to think them through may have had a higher chance to solve them than those who rushed or applied guessing strategies.

A tale of two test setups: The effect of randomizing a popular conceptual survey in electricity and magnetism

Two of the questions, Q13 and Q21, had medium difficulty but some of their answer-options may be eliminated via simple logic; reading and applying logic may take more time than the 13.5 seconds that the *quick*s spent on average on a question on WebAssign. The questions Q10, Q12, and Q27 are hard but again logical reasoning may have had a better chance for correct than the rushing-through strategy. At this stage of the data analysis, we couldn't find possible explanations for the differences in popularity due to the time-in-testing for the questions Q6, Q25, Q26, and Q31

For four BEMA-questions (Q7, Q8, Q9 and Q11) with the options presented in randomized order on WebAssign, the *quick* and the *valid* popularities had comparable values at the 95% confidence level. We thought of the following interpretation. If the key is at the position A or B whether the two options are randomized or not, the chance to get the correct answer by *quick* guessing could equal the chance to get the correct answer after several minutes of naive thought. This was the case of the question Q8. If a hard question has many randomized options, on the one hand the key may land at A, B, or C to benefit the blind guessers, on the other hand the persons with little conceptual knowledge will hardly be able to tell one option from the other. Hence, fast and slow respondents in quartile 1, end up with nearly equal chances for a correct answer to the hard questions Q7, Q9, and Q11 (see Figure 4).

The same message of the equality of chances, to get a correct answer by either guessing or thinking but being distracted by the multitude of options, may hold for two questions presented with the options in fixed order on WebAssign: Q2 (easy) and the Q24 (hard). See Figure 4. The reader should be aware that the results and the speculative discussion in this subsection pertain to the case when the BEMA-questions are listed in fixed order all on the same page, i.e., on WebAssign.

a) Comparing item popularities for *quick* and *valid* time categories, for questions in either fixed or random order

In this section we refer to data collected on both WebAssign and the local platform. As mentioned earlier, the size effect offsets the effect of the test presentation for the quartiles 2 and 3, therefore we restricted the comparisons *quick* vs. *valid* to quartile 1 (low scores). We investigated if the order in which the questions were presented affected the item popularity or not. We calculated the popularities for the questions with the options listed in fixed order (question groups 3 and 4) separately for each type of presentation and time category.

Figure 5 illustrates the popularity calculations for quartile 1, for the two types of test presentations. The item-popularities of the questions presented in fixed order, all on the same page, are given on the abscissa. The popularities of the same questions presented in shuffled order, one question per page, are given on the ordinate. The shape and color of a point in this plot indicate the time category. The dashed arrows emphasize the most obvious popularity shifts. The differentiation *quick* vs. *valid* was amplified by shuffling the order of the questions, and the greatest effect occurred for the question Q1, the easiest in the BEMA test. In the shuffled presentation, question Q1 can land as far away as position 22 in the list of questions. Moving Q1 away from the top of the test decreased the probability of a correct answer to 3/4 for *valid*s but dramatically to 1/4 for *quick*s.

In addition, one can see significant differences between the answers given by the *quick*s and the answers given by the *valid*s to the questions Q1, Q3, and Q21 (already identified in Figure 4), and for the questions Q2 and Q6. We conclude again that the persons in the *quick*-category have a significantly different strategy to answer test questions than those in the *valid*-category.

A tale of two test setups: The effect of randomizing a popular conceptual survey in electricity and magnetism

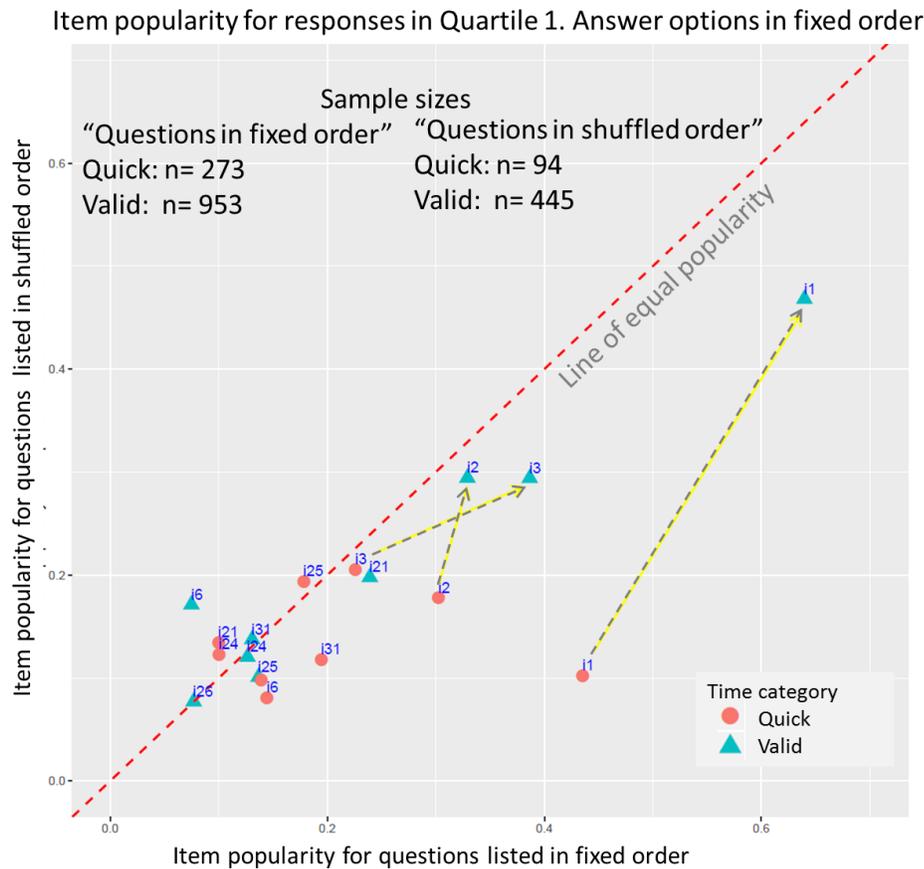

**Figure 5: Item popularities for two different test presentations.** The questions shown in this plot have the options listed in fixed order. Data from pre- and post-instruction are pooled together. The item popularities were calculated for quartile 1, separately for each of the two time categories. The shape and color of a point indicate the time category. The point's coordinates represent the item-popularities of the same question calculated 1) when questions were in fixed order and all on the same page (on the abscissa), and 2) when questions were in shuffled order, one question per page (on the ordinate). Those items on or close to the line of equal popularity had similar chances to be answered correctly independently of the test presentation. Dashed arrows indicate the most obvious differences between the popularity calculated for *quicks* and the popularity calculated for *valids*, and correspond to the questions Q1, Q2, and Q3.

b)  Comparing the option choice frequencies for questions with options in randomized order

Next we calculated the frequency of choosing an option [20] if the option position was randomized, for question-groups 1 and 2 and for respondents from quartile 1, when the questions were listed in fixed order (data collected on WebAssign). As we stated earlier, if the options come in randomized order the answer on a position in the list, e.g. option A, could be any of the answers listed by the authors for that question. We found a significant difference in the choices made by the *quick*s and *valid*s for just two questions, both of them we classify as being "hard", Q10 and Q27. The significance was checked via the chi-square test at the 0.001 alpha-level. The results are listed in [18] Appendix 5.

Table 1-4 shows the distribution of the answer choices for the two questions for which the option choices depended significantly on time-category. The last option reads "none of the above" for both of these questions and stayed fixed on the last position of the list during the option randomization. Question Q10 has eight very similar symbolic answer-options. The *quick* responders preferentially chose one of the first four options while the choices of the *valid* respondents were spread almost uniformly, mirroring the option randomization. The first seven options of the question Q27 are short similar symbolic formulae. Among these, two are strong distracters and one is correct. The fact that the frequency distribution for *valid* responders in Table 3(b) has three peaks may indicate the presence of the strong distracters.

Pattern analysis revealed systematic guessing for quartile 1 and short time-in-testing, supporting the hypotheses "the responses corresponding to short test durations are most likely guessed". For twelve questions in BEMA, we found significant differences between the answers of the *quick* test takers and of the *valid*-time responders, with all other parameters other than time being controlled. We could not detect significant differences between *quick* and *valid* answers for six BEMA questions in this study. Guessing as a confounding factor intervened in same extent in the regular *valid* time responses, i.e., in the answers to question Q8 given by the persons in quartile 1.

**Table 1-4 – Option choice frequency for two hard questions with randomized options.** Responses submitted on WebAssign for questions listed in fixed order. Data from pre- and post-instruction pooled together (a) for question Q10 and (b) for question Q27.

(a)

| Question Q10 *(hard)* | | Quartile 1 | Option | | | | | | | | | |
|---|---|---|---|---|---|---|---|---|---|---|---|---|
| Time-category | Sample size | | A | B | C | D | E | F | G | H | I | Sums (%) |
| *quick* | 269 | Proportions (%) | 19 | 19 | 18 | 15 | 9 | 9 | 4 | 4 | 3 | 100 |
| *valid* | 952 | | 12 | 14 | 12 | 14 | 12 | 12 | 11 | 11 | 1 | 100 |

(b)

| Question Q27 *(hard)* | | Quartile 1 | Option | | | | | | | | |
|---|---|---|---|---|---|---|---|---|---|---|---|
| Time-category | Sample size | | A | B | C | D | E | F | G | H | Sums (%) |
| *quick* | 269 | Proportions (%) | 13 | 18 | 21 | 24 | 9 | 9 | 4 | 2 | 100 |
| *valid* | 944 | | 16 | 13 | 18 | 17 | 15 | 10 | 8 | 2 | 100 |

**PART II. Building homogeneous subpopulations of persons and responses**

The overall goal of our study is to see if randomizing the questions, the options, or both, affects the results of the BEMA testing. If this study were to be conducted in the future, we would give the same questions to groups of persons extracted from the same population, one group randomly assigned to one type of question presentation. However, because our study is retrospective, we retained for analysis only the responses given by homogeneous subpopulations.

The full population in testing at TAMU was not homogeneous (see Table 1-5). As discussed in detail in **PART I**, we identified different answering strategies based on time-in-testing; guessing was the main strategy of the *quick* responders, who scored mostly in quartile 1. We found unexplained features, which need further analysis, in the responses of the multi semester test takers. We also observed complex variations of the score and time-in-testing for those persons who took the BEMA three or four times during our study. A relevant example of six persons enrolled in the PHYS208 three times who took the BEMA test five times during our study is presented in [18] Appendix 6.

The study of the *quick* takers and multi takers, and the driving forces behind their answers, will be the subject of future journal papers. For the present analysis, we set apart the data from the "Quick" and "Multiple" responses that could introduce confounding variables into our study. As one would expect, cleaning the data of *quick* and multiple responses had a significant effect on the mean scores and on the person distribution in the score quartiles. The interested reader can find the comparative score statistics in [18] Appendix 8.

**Table 1-5–Categories of responses collected in our study.** The persons in the category "repeat" have two responses collected during the same semester, one in the bin PRErepeat and the other in the bin POSTrepeat. The persons in the category "multiple" submitted three, four, or five responses at different dates of testing. The responses in the time categories "*quick*" and "*valid*" followed significantly different patterns. "Subpopulation PRE" and "Subpopulation POST" are two homogeneous groups of responses.

| | PREonly | PRErepeat | POSTrepeat | POSTonly | Multiple |
|---|---|---|---|---|---|
| Quick | *69* | *129* | *202* | *2* | *60* |
| Valid | *977* | *2059* | *1986* | *66* | *527* |
| | | Subpopulation PRE | Subpopulation POST | | |

After separating off the responses in the categories "*Quick*" and "Multiple", we compared the responses in the groups "PREonly" and "PRErepeat". No significant difference in average score, average time, average rate of answering, and option choice distributions was found. We concluded that the subpopulation of 3036 *valid* responses, labeled PRE, was homogeneous. Similarly, we found no significant difference between the groups of responses POSTonly and POSTrepeat. Hence, we had a second homogeneous subpopulation of 2052 *valid* responses labeled POST.

**PART III. The effect of the test presentation on scores**

With these two homogeneous subpopulations of responses, we calculated the item popularities separately for PRE and POST, for each type of test presentation, and checked for significance in the difference in popularity via normal approximation testing at 0.001-level. The results for those questions perceived differently by respondents within the same range of scores (within the same score quartile) are listed in Table 1-7 and 1-6, and the results of the item popularity for the overall subpopulations PRE and POST are plotted in Figure 6 and Figure 7.

At pre-instruction, four questions from Group 3 (Q1, Q2, Q3, and Q21) (see **Table 1-2**) were perceived to have significantly different item popularity if the test presentation was changed. These questions had the order of the options fixed and the correct answer at A or B, for both test presentations that we compared. At post-instruction, only the questions Q1, Q7, and Q12 were perceived to have significantly different item popularity if the test presentation was changed. Question 1 had the options in fixed order in both test presentations. The questions Q7 and Q12 had the options randomized on WebAssign; these questions addressed a high level of conceptualization and their answer-options were verbose. The difference in popularity for these two questions came from the quartiles 3 and 4, therefore from groups of respondents who most likely read carefully through the question before submitting an answer.

The trio of dependent questions Q1, Q2, and Q3 were perceived to have different popularity when they were separated, and mixed with other questions on the local TAMU server. Question Q1 had significantly less correct answers if it was presented somewhere in the middle of the question list than if it was at the top of the list. Its popularity for the quartile 1 at pre-instruction was 0.652 if presented as the first question but 0.502 if presented in a random position (a 15% difference). Also for quartile 1, at post-instruction, Q1's popularity was 0.596 if presented first in the list but 0.365 if presented in a random position (a 23% difference). For the other score-quartiles at either pre- or post-instruction the popularity differences for question Q1 were under 5%. We inferred that the test-takers with low scores who answered question Q1 at the beginning of the test had higher chances to get a correct answer than if this question appeared in the middle of the test.

The questions Q2 and Q3 received significantly more incorrect answers if they were presented as a continuation of the question Q1 than if they were presented as separate questions with their own diagrams. The trio of questions Q1, Q2, and Q3 was designed to be graded as dependent on each other. This design probably introduces a halo effect of Q1[12] over its two followers due to the partial credit grading employed on the WebAssign platform. The halo effect vanished when these three questions were presented on separate pages and not in sequence on the local platform.

When question Q3 was presented in shuffled order, it was graded for correctness only (independent answers model). If question Q3 was presented fixed in the trio (Q1, Q2, Q3), it was graded for answer consistency (partial credit model). An incorrect answer for question Q3 received one point if it was identical with the answer for the precedent Q2. At pre-instruction on WebAssign, out of the answers for Q3 graded with "1", 408 were incorrect and 1036 were correct; 121 correct answers lacked consistency and were graded with "0". In Figure 6 , question Q3 appears below the line of equal popularity because the bonus for consistency was given on WebAssign, but not on the local platform.  For a fair comparison, we regraded the answers for Q3 from WebAssign in the independent model. Table 1-6 gives the popularity at pre-instruction for the two different scoring methods and two different positions of Q3 in the question list. If Q3 was graded only for correctness, its popularity at pre-instruction was 0.558 when presented in trio, but 0.649 when presented detached from Q1 and Q2.  The interested reader can find the option choices of the subpopulation PRE when answering Q3 in [18] Appendix 9.

**Table 1-6 – Popularity calculations for Q3, for subpopulation PRE and two scoring methods**. Cleaned data.

| Question presentation | Scoring | |
| --- | --- | --- |
| | Partial credit popularity | Independent popularity |
| in  trio with Q1 and Q2 (on WebAssign) | 0.696 | 0.558 |
| independent item (on the local platform) | NA | 0.649 |

Question 21, which refers to a bar magnet, received significantly less correct answers when it was presented separately from the other questions pertaining to magnetism at pre-instruction; no difference was found at post-instruction. Question Q21 addresses a concept probably unfamiliar to students at pre-instruction, magnetic field.  It was identified in a previous research [11] as having a significant positive association  with the question Q22 (not included in this study) and a significant negative association with three other BEMA questions. Hence we suspect that Q21 most likely belongs to a group of dependent questions.

In order to understand why the scores for Q21 were affected by shuffling the order of the questions at pre-instruction, we studied the responses to the similarly constructed question Q6. Q6 refers to the deviation of a charged particle in an electric field. Q21 and Q6 have almost identical answer-options consisting of vector diagrams and verbal statements; both have a strong distracter. The overall popularity of Q6 was the same no matter the position of this question in the test. The popularity of Q21, however, was lower when it was separated from the other questions on magnetism Q20, Q22, and Q23. Did the popularity change for Q21 but not for Q6 because the test-takers at pre-instruction were less familiar with magnetic fields than with electric fields? The interaction effect between familiarity with the concept and the shuffling of the questions' order needs further investigation.

---

[12] Coulomb's law suffices for solving Q1, but Newton's 3rd law must be recalled before solving Q2 and Q3.

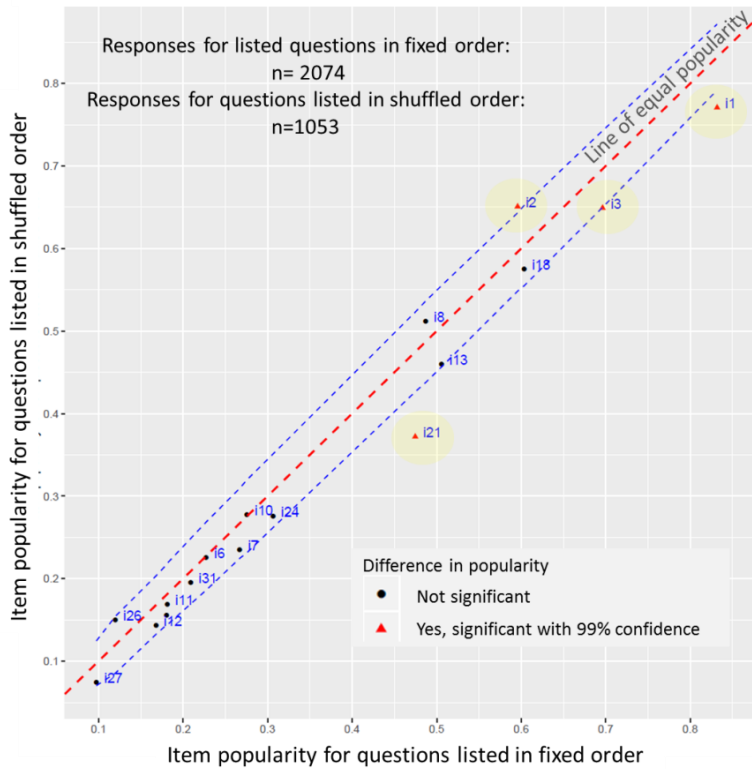

**Figure 6 -Graphical comparison of the item popularities for two test presentations, at pre-instruction. R**esponses from sample PRE (cleaned data). The dashed lines represent the 99% confidence interval for the difference in popularity. The highlighting circles emphasize the questions of significantly different popularities in the two test presentations: Q1, Q2, Q3, and Q21.

Item popularities for responses at post-instruction in two test setups

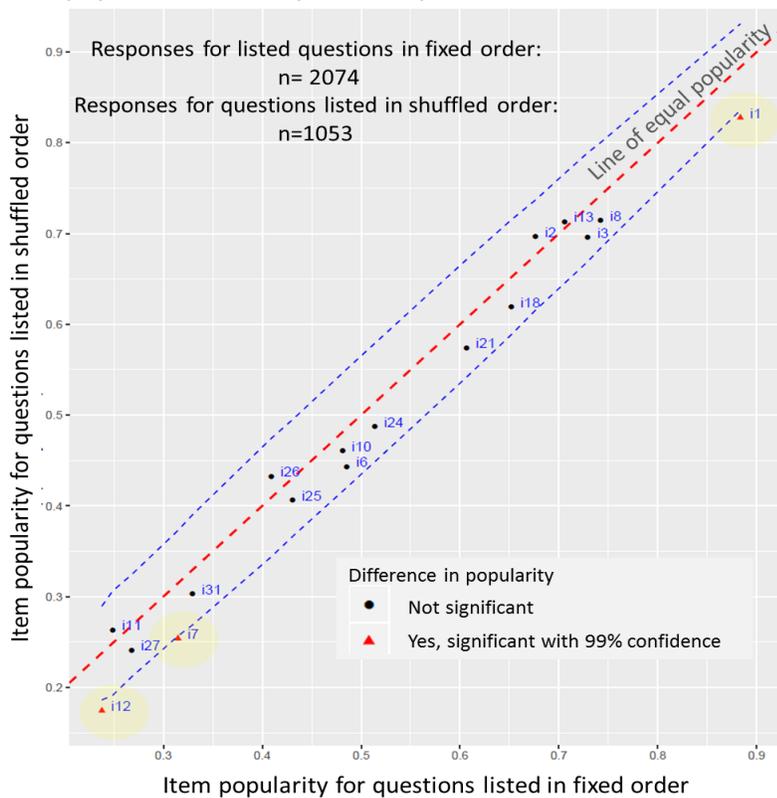

**Figure 7** -Graphical comparison of the item popularities for two test presentations, at post-instruction. Responses from sample POST (cleaned data). The dashed lines represent the 99% confidence interval for the difference in popularity. The highlighting circles emphasize the questions of significantly different popularities in the two test presentations: Q1, Q7, and Q12.

**Table 1-7 – Questions that changed their popularity significantly when they were presented in shuffled order, at pre-instruction.** Comparisons between groups of responses corresponding to the same score range (quartile), sample PRE (cleaned data). A zero popularity difference signifies no effect of the test presentation on the BEMA questions. We found no significant popularity differences for the question-group 1, sample PRE.

| Question group | Question name | Question difficulty | Score quartile | Item popularity | | Popularity difference Limits of the 99% CI | |
|---|---|---|---|---|---|---|---|
| | | | | WebAssign | Local | Lower | Upper |
| 2 | Q13 | medium | 1 | 0.325 | 0.230 | 0.018 | 0.172 |
| | | | 3 | 0.688 | 0.572 | 0.022 | 0.209 |
| | Q7 | hard | 2 | 0.263 | 0.169 | 0.002 | 0.146 |
| | Q11 | hard | 1 | 0.070 | 0.131 | -0.117 | -0.005 |
| | Q27 | hard | 3 | 0.121 | 0.049 | 0.020 | 0.0122 |
| 3 | Q1 | easy | 1 | 0.652 | 0.502 | 0.063 | 0.237 |
| | | | 3 | 0.970 | 0.923 | 0.002 | 0.093 |
| | Q3 | easy | 1 | 0.418 | 0.286 | 0.050 | 0.215 |
| | | | 3 | 0.928 | 0.867 | 0.001 | 0.121 |
| | Q21 | medium | 2 | 0.488 | 0.306 | 0.098 | 0.266 |
| | | | 3 | 0.616 | 0.456 | 0.064 | 0.255 |
| | | | 4 | 0.893 | 0.724 | 0.053 | 0.286 |
| | Q24 | hard | 2 | 0.281 | 0.202 | 0.005 | 0.152 |
| | | | 3 | 0.469 | 0.344 | 0.032 | 0.219 |
| 4 | Q6 | hard | 1 | 0.064 | 0.176 | -0.173 | -0.051 |
| | | | 3 | 0.357 | 0.235 | 0.035 | 0.208 |
| | | | 4 | 0.725 | 0.433 | 0.152 | 0.432 |

A tale of two test setups: The effect of randomizing a popular conceptual survey in electricity and magnetism

We found a significant change in popularity if the order of the questions was shuffled for Q7 and Q12. These two questions in "Group 2" had the order of the options randomized on WebAssign, but fixed on the local platform. They were hard questions and they were answered mostly incorrectly at pre-instruction. The change in popularity was observed at post-instruction and came mostly from the high scoring test-takers in quartiles 3 and 4. Seeking an explanation for the change in popularity, we will analyze the option choices for these two questions in PART IV.

**Table 1-8 – Questions that changed their popularity significantly when they were presented in shuffled order at post-instruction.** Comparisons between groups of responses corresponding to the same score range (quartile), sample POST (cleaned data). A zero popularity difference signifies no effect of the test presentation on the BEMA questions.

| Question group | Question name | Question difficulty | Score quartile | Item popularity | | Popularity difference Limits of the 99% CI | |
|---|---|---|---|---|---|---|---|
| | | | | WebAssign | Local | Lower | Upper |
| 1 | Q8 | easy | 3 | 0.756 | 0.626 | 0.009 | 0.251 |
| 2 | Q7 | hard | 3 | 0.328 | 0.167 | 0.052 | 0.270 |
| | Q10 | hard | 3 | 0.472 | 0.309 | 0.037 | 0.288 |
| | Q12 | hard | 4 | 0.376 | 0.224 | 0.070 | 0.235 |
| 3 | Q1 | easy | 1 | 0.596 | 0.365 | 0.045 | 0.418 |
| 4 | Q6 | hard | 4 | 0.758 | 0.637 | 0.032 | 0.209 |

**PART IV.  The effect of the test presentation on the answer choices**

We compared the option choice frequencies on WebAssign and on the local platform across the four score-quartiles (see Figure 8, Figure 9, and Figure 10). To make this comparison we calculated these frequencies for each of the two homogeneous subpopulations PRE and POST, and for each type of question presentation. For the questions with options listed in fixed order (the question groups 3 and 4) we found that the option choice frequencies changed significantly if the test presentation changed only for the responses in quartile 1 (total scores from 0 to 7) and only for the questions Q1 and Q2. We observed a significantly different trend of choosing an option from a fixed list in the case when these two questions were listed together at the top of the same page, and shared the same picture in the test, as opposed to the case when these questions were listed away from each other, on different pages, and possibly in an inverse sequence.

For question Q1 and respondents with low scores from quartile 1, Figure 8a), the option E that signals no knowledge of Coulomb's law was more popular if question Q1 was on top of the questions' list then somewhere in the middle of the list.  The percentage of correct answers out of the total submitted for the question Q1 was the highest if Q1 was on the top of the questions' list.

Coulomb's law suffices for solving Q1, but Newton's 3$^{rd}$ law must be recalled before solving Q2 and Q3. For question Q2 in Figure 8 b), one sees that the option E, which goes against any understanding of the Coulomb's law, was the preferred answer if Q2 was presented at the top of the list immediately after the question Q1. The option F was another distracter chosen by quartile 1 at pre-instruction. The answer-option F assumes that the smaller charge exerts a smaller force and contradicts Newton's 3$^{rd}$ law.

116 test takers, representing a 5.6% of the total at pre-instruction on WebAssign, chose option E for both questions Q1 and Q2 when these two questions were presented in sequence; 89 of them had the scores in the first quartile (a 26% of quartile 1).  By comparison, 30 test takers representing a 2.8% of the total at pre-instruction on the local platform chose option E for both Q1 and Q2 when these two questions where not in sequence. At post-instruction, the proportion of test takers in quartile 1 choosing option E for both Q1 and Q2 was of 10.3% when these questions appeared in sequence and of 4.2% when they were mixed among other questions. We may infer that presenting the questions Q1 and Q2 sequentially doubled the chance that they will get the same wrong answer from someone who mastered neither Coulomb's law, nor Newton's 3rd law.

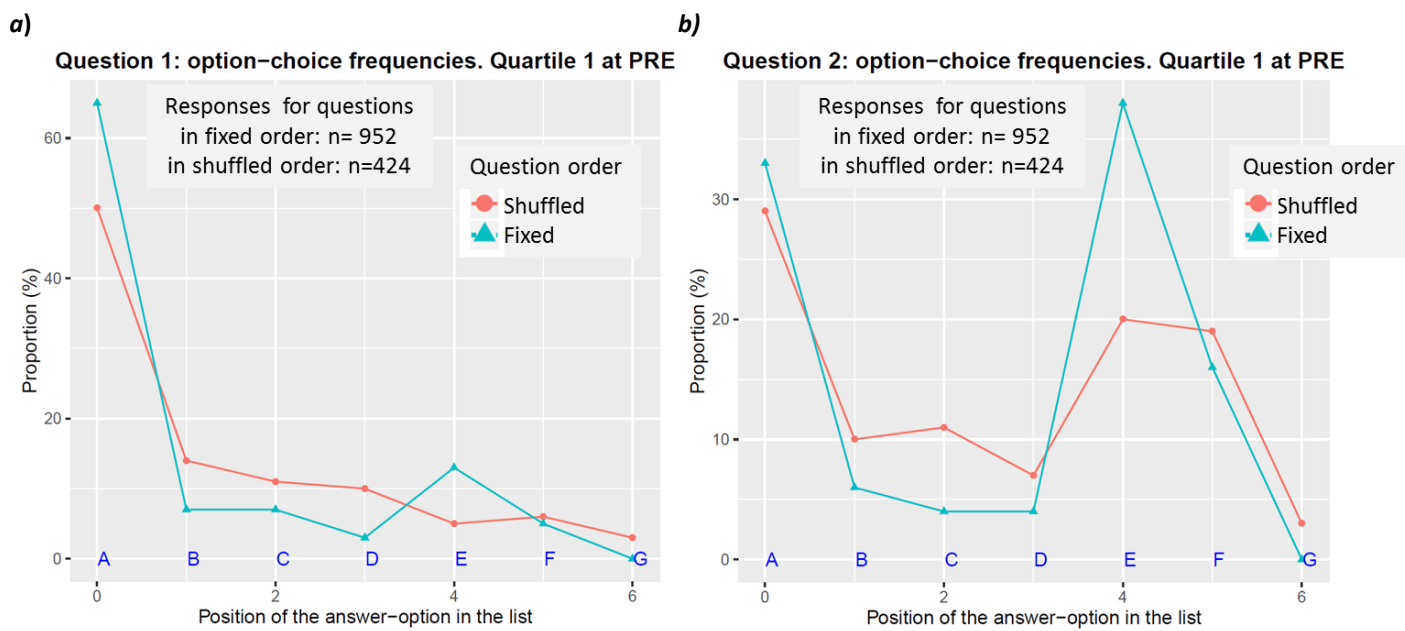

**Figure 8 – Option choice frequencies for two BEMA questions with the options listed in fixed order.** Proportions calculated for two different test presentations, questions listed in fixed order versus questions listed in random order, sample PRE and score quartile 1 (cleaned data). Based on the Chi-Square test, the distribution of the option choices is significantly affected by the type of presentation for the questions Q1 and Q2.

Question 21, with the options in fixed order in both test presentations, came out with significantly different popularities at pre-instruction. The option choice frequencies plots in Figure 9 show that this difference came mostly from the choice of the distracter options E (which represents the symmetrical direction to that of A) if the order of the questions was shuffled. No significant difference in the option choices for Q21 was observed at post-instruction.

We found statistically significant overall differences in the popularities if the options were randomized on WebAssign for only two questions, Q7 and Q12; the differences were stronger at post-instruction. We show the option choice frequencies of these questions, for quartile 4 at post-instruction, in Figure 10. The option choice distribution is flattened if the order of the options was randomized, as we expected. In the case of the fixed order of the options for question Q7, we noticed two very strong distracters: "not be affected by the charges on the wall since rubber is an insulator." at position A and "The field must be non-zero because the flowing current produces an electric field" at position D.

Some of the high scoring students chose a distracter positioned on the option-list before the correct answer for Q7 and Q12. We may interpret this as follows. When the order of the questions was shuffled, with each question appearing on a separate page and with no conceptual continuity from one page to another, the attention effort necessary for evaluating the answer-options was much greater than when the questions were grouped on concepts and presented on the same page. Both the questions Q7 and Q12 have verbose answer-options that require careful evaluation. We believe that a part of the test-takers stopped reading these long verbose options somewhere in their middle, maybe not reading all the options in the list.

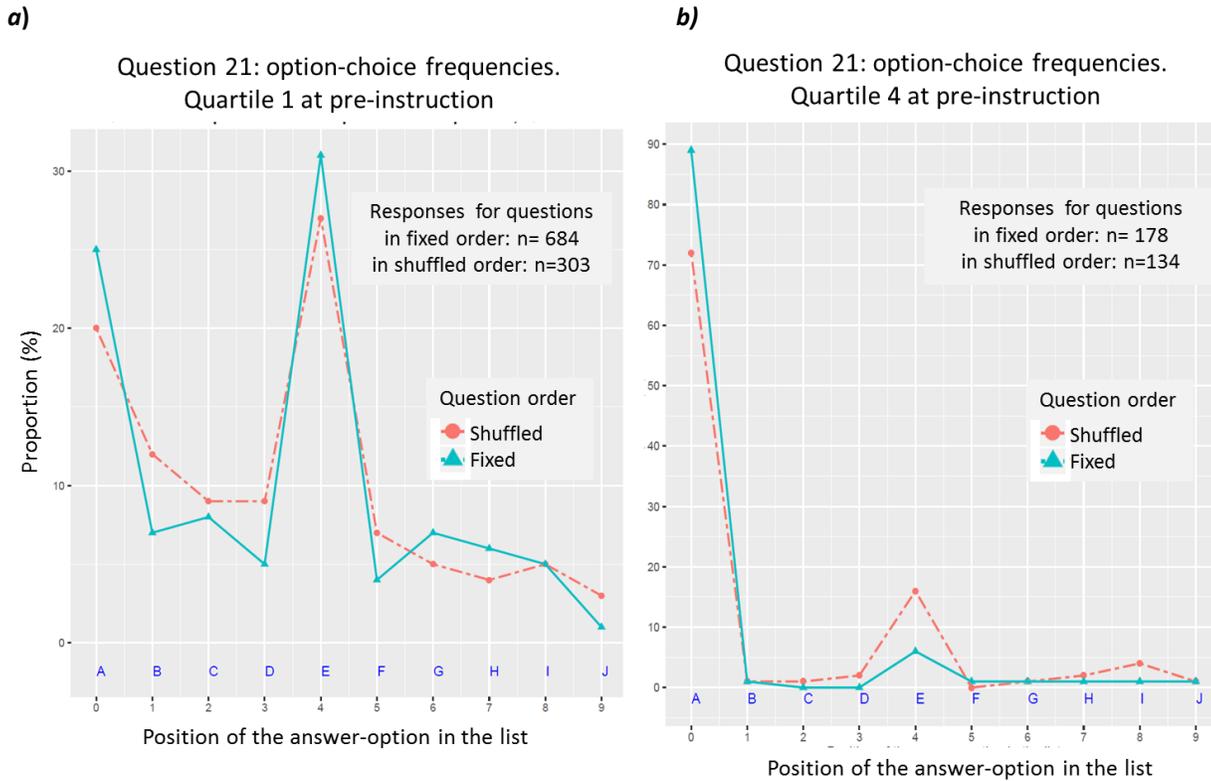

**Figure 9 – Option choice frequencies for the BEMA question Q21 with options listed in fixed order.** Proportions calculated for two different test presentations: questions in fixed order and questions in random order, for the sample PRE and the score quartiles 1 and 4 (cleaned data). Based on the Chi-Square test results, the distribution of the option choices is significantly affected by the type of presentation for the sample PRE.

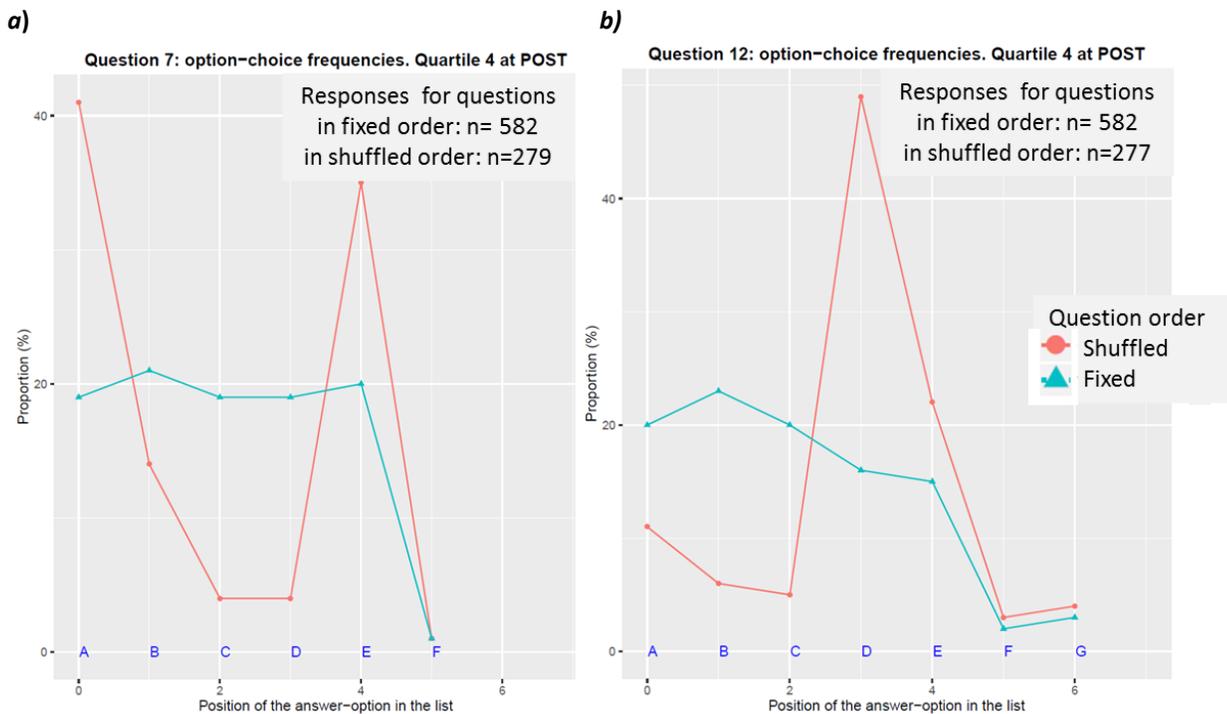

**Figure 10 – Option choice frequencies for the BEMA questions Q7 and Q12 with randomized options on WebAssign.** Proportions calculated for two different test presentations, questions in fixed order and questions in random order, sample POST and score quartile 4 (cleaned data). Based on Chi-Square test results, the distribution of the option choices is significantly affected by the type of presentation.

We attributed the significant decrease in the proportion of correct answers for the questions Q7 and Q12 to the shuffled of the question list and not to the option randomization on WebAssign. We had a confirmation of this interpretation from two interviews with students repeating the pre- and post-instruction BEMA testing on the local platform, and making A-s in their PHYS208 class. We have not conducted systematic interviews with the students taking the BEMA testing on WebAssign.

### F.   Conclusions

We described a retrospective study of the responses to the BEMA collected from a large population of students from a large public university using two different testing setups. It is no coincidence that the title of our study is a paraphrase of the 2009 paper[4] that compared the average BEMA scores across two different curricula. After one decade, we are telling another story about the BEMA. This time the focus is not on the learning gains estimated via BEMA, but on the interpretation of the data collected from the BEMA testing itself.

Does the randomization of the answer-options for nine BEMA questions, on WebAssign, affect the scores for the BEMA test? We found that the answer is "yes" for the case of the fast rate answer submissions (less than 13.5 seconds on average per answer) and generally "no" for the test-takers spending an average of more than 13.5 seconds per answer. Our study did not detect any significant effect of the option randomization alone on the BEMA scores, for *valid* times-in-testing.

It appeared that option randomization is not necessary to counterbalance guessing, if the *quick* responses are separated from the pool, and if there are no breaches in the test security. We believe that the findings related to fast answering rates apply not only to the BEMA but to all the multiple choice tests, and that the time-in-testing should always be used as a validation criterion for computerized testing. Another response validation criterion that we suggest be implemented is pattern analysis, which can separate the obvious guessing strategies of the test takers.

The BEMA test with the questions in fixed order follows the typical concept sequence of an E&M course. However, shuffling the questions produces unusual sequences of interwoven concepts. Does the shuffling of the questions affect the scores of the BEMA test?  Our study found that presenting the questions in shuffled order, when keeping the position of the answer-options fixed, affected the answers to four questions at pre-instruction. Three of these questions address the same concept, Coulomb's law, and were written to depend on each other. Q1, presented as an independent question somewhere in the middle of the test, received a smaller percentage of correct answers than if it was the first question of the test. Q2 and Q3 received greater percentages of correct answers, at pre-instruction, when they were presented separately than if they followed immediately after Q1 and sharing the diagram with it. Finally, the question Q21 received a smaller percentage of correct answers at pre-instruction when it was presented separately from the group of questions covering magnetic fields than if it was presented with the group. The groups of dependent questions were answered differently if they were presented as separate, independent questions.

Our conclusion is that shuffling the independent questions does not affect the scores of the BEMA test, while shuffling the dependent questions does affect the scores. Hence, in order to make comparisons across instructors, the groups of dependent questions in BEMA must be presented in the original sequence, and graded in the partial credit model.

We have detected a combined effect of the option randomization and of the shuffling of the questions, at post-instruction, in the case of two questions with long, verbose answer-options. We attributed the significant decrease in the proportion of correct answers for the questions Q7 and Q12 to the shuffled question list and not to the option randomization. This combined effect needs further investigation.

Can the mean score from the BEMA testing be employed to compare samples of test-takers from different instructors, even different curricula? According to our study, the answer is "no", if the *quick* responses and those corresponding to blind guessing patterns are included in the calculation of the mean score. The mean score of all responses collected on WebAssign in our study was 10.62; after dropping the quick and multiple submissions, the clean mean score was 11.79. The greatest difference came from the quartile 1. Thus, we believe that the data collected need to be *cleaned* before performing statistical analysis and results reporting.

Even if the data are cleaned, the mean score is not a relevant statistic for the mostly non normal score distributions that we usually obtain from pre-post testing. As long as the BEMA questions are not invariantly calibrated, we suggest comparing the proportions of cleaned responses in the relevant score intervals (i.e., for our data these were quartiles) instead of comparing the overall mean score. This allows one to compare the differences in the shapes of these distributions across instructors/curricula/test presentations.

## *Acknowledgements*

This study would not have been possible without the involvement of Tatiana Erukhimova, Alexey Belyanin, and of the teaching assistants who helped organize and supervise the online testing at TAMU during the 2014-2017 academic years. We also thank to Anthony Cooper and Shuya Ota for importing, merging, and coding the BEMA data collected from the online platforms. Last but not least, we wish to thank Bruce Sherwood, Ruth Chabay, and Sam McKagan for the discussion on the possible effects of the answer-options randomization on WebAssign.

# Supplemental Materials

**Appendix 1 - Item popularities calculated for the pooled 6077 responses in our study.**

The counts for answers not submitted (the NA responses) **are** not included.

| item | quart | count.0 | count.1 | popularity (%) | difficulty |
|------|-------|---------|---------|----------------|------------|
| i1 | 1 | 776 | 920 | 54.2 | easy |
| i1 | 2 | 265 | 1336 | 83.4 | easy |
| i1 | 3 | 105 | 1245 | 92.2 | easy |
| i1 | 4 | 21 | 1217 | 98.3 | easy |
| i1 | Sum | 1167 | 4718 | 80.2 | easy |
| i10 | 1 | 1427 | 267 | 15.8 | hard |
| i10 | 2 | 1209 | 391 | 24.4 | hard |
| i10 | 3 | 868 | 482 | 35.7 | hard |
| i10 | 4 | 455 | 786 | 63.3 | hard |
| i10 | Sum | 3959 | 1926 | 32.7 | hard |
| i11 | 1 | 1530 | 158 | 9.4 | hard |
| i11 | 2 | 1349 | 252 | 15.7 | hard |
| i11 | 3 | 1055 | 298 | 22 | hard |
| i11 | 4 | 789 | 444 | 36 | hard |
| i11 | Sum | 4723 | 1152 | 19.6 | hard |
| i12 | 1 | 1475 | 215 | 12.7 | hard |
| i12 | 2 | 1344 | 252 | 15.8 | hard |
| i12 | 3 | 1139 | 205 | 15.3 | hard |
| i12 | 4 | 864 | 379 | 30.5 | hard |
| i12 | Sum | 4822 | 1051 | 17.9 | hard |
| i13 | 1 | 1222 | 460 | 27.3 | medium |
| i13 | 2 | 822 | 771 | 48.4 | medium |
| i13 | 3 | 439 | 905 | 67.3 | medium |
| i13 | 4 | 161 | 1067 | 86.9 | medium |
| i13 | Sum | 2644 | 3203 | 54.8 | medium |
| i18 | 1 | 841 | 837 | 49.9 | easy |
| i18 | 2 | 631 | 970 | 60.6 | easy |
| i18 | 3 | 520 | 826 | 61.4 | easy |
| i18 | 4 | 386 | 859 | 69 | easy |
| i18 | Sum | 2378 | 3492 | 59.5 | easy |
| i2 | 1 | 1167 | 524 | 31 | easy |
| i2 | 2 | 656 | 947 | 59.1 | easy |
| i2 | 3 | 337 | 1018 | 75.1 | easy |
| i2 | 4 | 112 | 1117 | 90.9 | easy |
| i2 | Sum | 2272 | 3606 | 61.3 | easy |
| i21 | 1 | 1342 | 341 | 20.3 | medium |
| i21 | 2 | 958 | 636 | 39.9 | medium |
| i21 | 3 | 572 | 774 | 57.5 | medium |
| i21 | 4 | 209 | 1028 | 83.1 | medium |
| i21 | Sum | 3081 | 2779 | 47.4 | medium |
| i24 | 1 | 1480 | 203 | 12.1 | hard |
| i24 | 2 | 1159 | 440 | 27.5 | hard |
| i24 | 3 | 770 | 578 | 42.9 | hard |
| i24 | 4 | 376 | 867 | 69.8 | hard |
| i24 | Sum | 3785 | 2088 | 35.6 | hard |
| i25 | 1 | 1451 | 231 | 13.7 | hard |
| i25 | 2 | 1314 | 283 | 17.7 | hard |
| i25 | 3 | 1021 | 324 | 24.1 | hard |
| i25 | 4 | 571 | 669 | 54 | hard |
| i25 | Sum | 4357 | 1507 | 25.7 | hard |
| i26 | 1 | 1470 | 140 | 8.7 | hard |
| i26 | 2 | 1370 | 191 | 12.2 | hard |
| i26 | 3 | 1054 | 265 | 20.1 | hard |
| i26 | 4 | 543 | 697 | 56.2 | hard |
| i26 | Sum | 4437 | 1293 | 22.6 | hard |
| i27 | 1 | 1555 | 128 | 7.6 | hard |
| i27 | 2 | 1436 | 158 | 9.9 | hard |
| i27 | 3 | 1213 | 141 | 10.4 | hard |
| i27 | 4 | 810 | 430 | 34.7 | hard |
| i27 | Sum | 5014 | 857 | 14.6 | hard |
| i3 | 1 | 1129 | 559 | 33.1 | easy |
| i3 | 2 | 546 | 1058 | 66 | easy |
| i3 | 3 | 238 | 1111 | 82.4 | easy |
| i3 | 4 | 96 | 1140 | 92.2 | easy |
| i3 | Sum | 2009 | 3868 | 65.8 | easy |
| i31 | 1 | 1423 | 235 | 14.2 | hard |
| i31 | 2 | 1268 | 309 | 19.6 | hard |
| i31 | 3 | 979 | 341 | 25.8 | hard |
| i31 | 4 | 715 | 497 | 41 | hard |
| i31 | Sum | 4385 | 1382 | 24 | hard |
| i6 | 1 | 1508 | 184 | 10.9 | hard |
| i6 | 2 | 1303 | 294 | 18.4 | hard |
| i6 | 3 | 916 | 437 | 32.3 | hard |
| i6 | 4 | 401 | 835 | 67.6 | hard |
| i6 | Sum | 4128 | 1750 | 29.8 | hard |
| i7 | 1 | 1426 | 264 | 15.6 | hard |
| i7 | 2 | 1224 | 378 | 23.6 | hard |
| i7 | 3 | 903 | 441 | 32.8 | hard |
| i7 | 4 | 757 | 487 | 39.1 | hard |
| i7 | Sum | 4310 | 1570 | 26.7 | hard |
| i8 | 1 | 1060 | 622 | 37 | easy |
| i8 | 2 | 768 | 832 | 52 | easy |
| i8 | 3 | 479 | 878 | 64.7 | easy |
| i8 | 4 | 192 | 1050 | 84.5 | easy |
| i8 | Sum | 2499 | 3382 | 57.5 | easy |
| i9 | 1 | 981 | 240 | 19.7 | hard |
| i9 | 2 | 854 | 314 | 26.9 | hard |
| i9 | 3 | 585 | 318 | 35.2 | hard |
| i9 | 4 | 468 | 337 | 41.9 | hard |
| i9 | Sum | 2888 | 1209 | 29.5 | hard |
| Sum | 1 | 23263 | 6528 | 21.9 | |
| Sum | 2 | 18476 | 9812 | 34.7 | |
| Sum | 3 | 13193 | 10587 | 44.5 | |
| Sum | 4 | 7926 | 13906 | 63.7 | |
| Sum | Sum | 62858 | 40833 | 39.4 | |

**Appendix 2 –Distribution of *quick* and *valid* respondents' raw data (test submissions)**

| Place | Quartile | Time-category | Counts |
|---|---|---|---|
| Local platform | 1 | *quick* | 94 |
| WebAssign | 1 | *quick* | 272 |
| Local platform | 2 | *quick* | 26 |
| WebAssign | 2 | *quick* | 53 |
| Local platform | 3 | *quick* | 5 |
| WebAssign | 3 | *quick* | 12 |
| Local platform | 1 | *valid* | 445 |
| WebAssign | 1 | *valid* | 952 |
| Local platform | 2 | valid | 427 |
| WebAssign | 2 | valid | 1118 |
| Local platform | 3 | valid | 479 |
| WebAssign | 3 | valid | 892 |
| Local platform | 4 | valid | 494 |
| WebAssign | 4 | valid | 808 |
| | | Total | 6077 |

**Appendix 3 -The answering pattern of a student who took our PHYS208 course in two subsequent semesters.** The answers submitted in less than 7min by the student Kim followed simple patterns. We show this case as evidence that answer selection independent of content happened in the group "*quick*".

| Semester | Spring16 | Spring16 | Summer16 | Summer16 |
|---|---|---|---|---|
| | PRE | POST† | PRE‡ | POST |
| time(min) | >7 | 4 | 2 | 2 |
| score | 5 | 8 | 2 | 6 |
| Question | Response | Response | Response‡ | Response |
| Q1 | A | A | G | G†† |
| Q2 | A | A | G | G†† |
| Q3 | B | A | I | I†† |
| Q6 | A | A | G | D |
| Q7 | A | A | F | D |
| Q8 | C†† | A | C | C†† |
| Q9 | E | A | E | D |
| Q10 | I†† | A | I†† | D |
| Q11 | I†† | A | I†† | D |
| Q12 | G†† | A | G†† | D |
| Q13 | E†† | A | E†† | D |
| Q18 | A | A | D†† | D |
| Q21 | C | A | J | D |
| Q24 | C | A | G | D |
| Q25 | A | A | H†† | D |
| Q26 | C | A | G | D |
| Q27 | B | A | H†† | D |
| Q31 | A | A | F | D |

†= all the responses in this column represent the first option in the list
‡=all the responses in this column represent the last option in the list
††=that option reads "none of the above" or "not enough information"

**Appendix 4 –Item popularities for *quick* and *valid* responses for the 18 questions in this analysis for data collected on WebAssign.**

| | Sample size | | | | | | | |
|---|---|---|---|---|---|---|---|---|
| | *quick* | *valid* | | Quartile 1, fixed order of questions | | | | |
| | 365 | 1398 | | Popularity difference | | | | |
| Question name | Difficulty | Question group | Item popularity | | Limits of the 95% CI | | | |
| | | | *quick* | *valid* | Lower | Upper | Significance |
| Q8 | easy | 1 | 0.387 | 0.351 | -0.019 | 0.092 | |
| Q9 | hard | 1 | 0.199 | 0.195 | -0.042 | 0.050 | |
| Q18 | easy | 1 | 0.353 | 0.539 | -0.241 | -0.130 | yes |
| Q7 | hard | 2 | 0.173 | 0.170 | -0.040 | 0.047 | |
| Q10 | hard | 2 | 0.070 | 0.176 | -0.138 | -0.072 | yes |
| Q11 | hard | 2 | 0.089 | 0.077 | -0.020 | 0.044 | |
| Q12 | hard | 2 | 0.159 | 0.119 | 0.000 | 0.082 | yes |
| Q13 | medium | 2 | 0.189 | 0.322 | -0.181 | -0.087 | yes |
| Q27 | hard | 2 | 0.104 | 0.071 | 0.000 | 0.067 | yes |
| Q1 | easy | 3 | 0.434 | 0.640 | -0.263 | -0.150 | yes |
| Q2 | easy | 3 | 0.301 | 0.327 | -0.079 | 0.027 | |
| Q3 | easy | 3 | 0.225 | 0.386 | -0.211 | -0.111 | yes |
| Q21 | medium | 3 | 0.100 | 0.240 | -0.178 | -0.102 | yes |
| Q24 | hard | 3 | 0.100 | 0.127 | -0.062 | 0.009 | |
| Q6 | hard | 4 | 0.144 | 0.075 | 0.030 | 0.108 | yes |
| Q25 | hard | 4 | 0.178 | 0.136 | 0.000 | 0.085 | yes |
| Q26 | hard | 4 | 0.139 | 0.077 | 0.024 | 0.100 | yes |
| Q31 | hard | 4 | 0.194 | 0.131 | 0.019 | 0.107 | yes |

**Appendix 5 - The effect of test presentation on the option choice frequency.**

The proportions of the option choices, for two types of question presentations, are compared via chi-squared tests at 0.01-level of significance. "same"= when question presentation changes, there is no significant change in the option choices; "change"- when question presentation changes, the option choices change significantly (99% cl). Results for responders in quartile 1.The number of answers submitted to a test question was generally smaller than the sample size.  Q9 was not included here since it was not assigned on the local platform.

Sample sizes in quartile 1

| Platform | Groups compared | |
| --- | --- | --- |
| | *Quick* | *Valid* |
| WebAssign | 273 | 953 |
| Local Platform | 94 | 445 |

| Question | | | Option choice | | p-value | |
| --- | --- | --- | --- | --- | --- | --- |
| **Name** | **Group** | Difficulty | *Quick* | *Valid* | *Quick* | *Valid* |
| Q18 | **1** | easy | *same* | *change* | 0.3498 | 0.0409 |
| Q8 | **1** | easy | *same* | *same* | 0.2997 | 0.6899 |
| Q10 | **2** | *hard* | *same* | *same* | 0.0842 | 0.7173 |
| Q11 | **2** | *hard* | *same* | *same* | 0.2264 | 0.5687 |
| Q12 | **2** | *hard* | *same* | *change* | 0.3199 | 0.0005 |
| Q13 | **2** | *medium* | *same* | *same* | 0.0704 | 0.1450 |
| Q27 | **2** | *hard* | *change* | *change* | 0.0001 | 3.10E-05 |
| Q7 | **2** | *hard* | *same* | *change* | 0.4850 | 0.0310 |
| Q1 | **3** | easy | *change* | *change* | 2.71E-05 | 0.0104 |
| Q2 | **3** | easy | *change* | *change* | 0.0001 | 0.0176 |
| Q3 | **3** | easy | *same* | *same* | 0.6983 | 0.6688 |
| Q21 | **3** | *medium* | *same* | *same* | 0.3431 | 0.9429 |
| Q24 | **3** | *hard* | *same* | *same* | 0.3642 | 0.8418 |
| Q6 | **3** | *hard* | *same* | *same* | 0.1539 | 0.3592 |
| Q25 | **4** | *hard* | *same* | *same* | 0.9562 | 0.9364 |
| Q26 | **4** | *hard* | *same* | *same* | 0.9878 | 0.9460 |
| Q31 | **4** | *hard* | *same* | *same* | 0.3201 | 0.9867 |

**Appendix 6 –** In the category of multiple test takers, **s**ix persons repeated the BEMA testing five times. Their responses were, with two exceptions, in the *valid*-time category and were spread over all four score-quartiles. These persons dropped the PHYS208 class during their first semester (no POST taken that semester). Their best score for BEMA was obtained at the end of their third semester in the same course. The persons' names are fictitious.  All multiple test taker data was cleaned from the sample that we are considering in this analysis.

| Person | Semester | Diagnostic | Time-cat | Score | Quartile |
|--------|----------|------------|----------|-------|----------|
| Babu   | fall14   | PRE        | v        | 5     | 1        |
|        | fall15   | PRE        | v        | 9     | 2        |
|        | fall15   | POST       | q        | 7     | 1        |
|        | spring15 | PRE        | v        | 7     | 1        |
|        | spring15 | POST       | v        | 18    | 4        |
| Casey  | fall14   | PRE        | v        | 4     | 1        |
|        | spring15 | PRE        | v        | 11    | 3        |
|        | spring15 | POST       | v        | 9     | 2        |
|        | summer15 | PRE        | v        | 9     | 2        |
|        | summer15 | POST       | v        | 15    | 4        |
| Mark   | fall14   | PRE        | v        | 8     | 2        |
|        | spring15 | PRE        | v        | 9     | 2        |
|        | spring15 | POST       | v        | 8     | 2        |
|        | spring16 | PRE        | v        | 13    | 3        |
|        | spring16 | POST       | v        | 13    | 3        |
| Wanda  | fall14   | PRE        | v        | 6     | 1        |
|        | summer16 | PRE        | v        | 5     | 1        |
|        | summer16 | POST       | v        | 10    | 2        |
|        | spring17 | PRE        | v        | 9     | 2        |
|        | spring17 | POST       | v        | 11    | 3        |
| Tod    | spring16 | PRE        | v        | 10    | 2        |
|        | summer16 | PRE        | v        | 9     | 2        |
|        | summer16 | POST       | q        | 4     | 1        |
|        | fall16   | PRE        | v        | 15    | 4        |
|        | fall16   | POST       | v        | 15    | 4        |
| Echo   | fall15   | PRE        | v        | 12    | 3        |
|        | summer16 | PRE        | v        | 14    | 3        |
|        | summer16 | POST       | v        | 10    | 2        |
|        | fall16   | PRE        | v        | 13    | 2        |
|        | fall16   | POST       | v        | 15    | 4        |

### Appendix 7 – Example of applying the Chi-square test of independence

We employed the Chi-square test for checking two hypotheses for independence: "the option choice frequencies are independent of the test presentation" for *valid* times-in-testing, and "the option choice frequencies are independent of the time-category" for questions presented in fixed order but having the options' order randomized.

We work here the example of comparing the option choice distributions for the BEMA-question Q1 presented in two different ways to our students from the sample PRE, quartile 1.  The reader should note that all students in sample PRE had *valid* times-in-testing and that question Q1 had the order of its options fixed in both test presentations.   This example is illustrated in Figure 8a) in the main text. The option choice frequencies for Q1, for the two test presentations, are shown in a cross-classification (contingency) table (**Error! Reference source not found.**). This contingency table has two rows (*i*=1, 2) and seven columns (*j*=1, 2,…, 7). The intersection of row *i* with column *j* represents the cell (i,j) of the table. The numerical value listed in a cell is an observed frequency $f_{ij}^o$ ( e.g., $f_{23}^o = 51$).

For computing the Chi-square statistic, one needs to first compute the marginal totals. Table 2-1 gives the column sums  $n_{+j}$ and the row ums $n_{i+}$ ( e,g, the sum on the column 3 is $n_{+3}$=113 and the sum on the row 2 is $n_{2+}$=391). The grand total representing the sample size is labeled *n*, where  *n*=1344 in this case.

The degrees of freedom of the Chi-square test are given by the number of independent cell counts, given that the sample size *n*, and the marginal totals $n_{+j}$ and $n_{i+}$ are fixed:

$$DF= (I*J-1) - (J-1) - (I-1) = (I-1)*(J-1)$$

In our case, J=7 and represents the number of answer-options for Q1; I=2 and represents the number of conditions (types of test presentations). Thus, DF=6.

**Table 2-1:  The observed option choice frequencies and the marginal sums for the BEMA-question Q1.** Question 1 was presented: (1) as the first question of the test and (2) somewhere in the middle of the test. Only the data from the sample PRE, quartile 1 are considered.

| Q1 *(easy)* | PRE , Quartile 1 | Option | | | | | | | |
|---|---|---|---|---|---|---|---|---|---|
| Question list | | **A** | **B** | **C** | **D** | **E** | **F** | **G** | Sums |
| Fixed (1) | | 610 | 60 | 62 | 40 | 123 | 57 | 1 | $n_{1+}$=953 |
| Shuffled (2) | Counts | 184 | 52 | 51 | 38 | 19 | 36 | 11 | $n_{2+}$=391 |
| | Sums | $n_{+1}$=794 | $n_{+2}$=112 | $n_{+3}$=113 | $n_{+4}$=78 | $n_{+5}$=142 | $n_{+6}$=93 | $n_{+7}$=12 | n=1344 |

The null hypothesis "the option choice frequencies are independent on the test presentation" was used to generate the expected homogeneous cell counts (frequencies) $f_{ij}^e$:

$$f_{ij}^e = (outcome_{frequency}) \cdot (weight_{condition}) \cdot (full\_sample_{size}) = \frac{n_{+j}}{n} \cdot \frac{n_{i+}}{n} \cdot n = \frac{n_{i+} \cdot n_{+j}}{n}$$

For example, the calculation of the expected frequency for the cell (i=2, j=3) in Table 2-1:

$$f_{23}^e = \frac{n_{+3}}{n} \cdot \frac{n_{2+}}{n} \cdot n = \frac{113 \cdot 391}{1344} = 32.87$$

A tale of two test setups: The effect of randomizing a popular conceptual survey in electricity and magnetism

The discrepancy between the counts observed and the counts expected under the null hypothesis is $\left(f_{ij}^o - f_{ij}^e\right)$. The usual Chi-square test uses the Pearson's residuals

$$\varepsilon_{ij} = \frac{\left(f_{ij}^o - f_{ij}^e\right)}{\sqrt{f_{ij}^e}}$$

The Chi-square statistic is calculated as

$$\chi^2 = \sum_{i=1}^{I} \sum_{j=1}^{J} \varepsilon_{ij}{}^2$$

For a rows-by-columns chi-square test, at least 80% of the cells must have an expected frequency of 5 or greater, and no cell may have an expected frequency smaller than 1. A correction for the situation when at least one cell has less than 5 counts was introduced by Yates for preventing the overestimation of the statistical significance. Yates's correction adjusts the expected counts with ½ unit closer to the observed value and the continuous Chi-square distribution closer to the discrete observed distribution [23]:

$$\varepsilon_{ij,Yates} = \frac{\left| f_{ij}^o - f_{ij}^e \right| - 0.5}{\sqrt{f_{ij}^e}}$$

The step-by-step calculations of the discrepancies, residuals, and sum of squares are shown in Table 2-2. Because one of the expected values was lower than 5, we employed the Yates's correction and obtained $\chi^2 = 95.44$. The probability to obtain such a large value $\chi^2 = 95.44$ purely by chance for DF=6 is lower than 0.0001, therefore the null hypothesis was rejected. The Chi-square test confirmed that the option choice distribution for the question Q1 depends significantly on the type of test presentation, for the quartile 1 of the sample PRE.

**Table 2-2 : Step-by-step calculations of the Chi-square statistic for the BEMA-question Q1 presented in two ways.** These data correspond to sample PRE, quartile 1.

| Question list | | | Option choice | | | | | | | Sums |
|---|---|---|---|---|---|---|---|---|---|---|
| | | | A | B | C | D | E | F | G | |
| Fixed | Counts | Observed | 610 | 60 | 62 | 40 | 123 | 57 | 1 | 953 |
| | | Expected | 563.01 | 79.42 | 80.13 | 55.31 | 100.69 | 65.94 | 8.51 | |
| | | Discrepancy | 46.99 | -19.42 | -18.13 | -15.31 | 22.31 | -8.94 | -7.51 | |
| | | Residual | 1.98 | -2.18 | -2.02 | -2.06 | 2.22 | -1.10 | -2.57 | |
| | | \|Yates Residual\| | 1.96 | 2.12 | 1.97 | 1.99 | 2.17 | 1.04 | 2.40 | |
| | | Squares | 3.84 | 4.51 | 3.88 | 3.96 | 4.72 | 1.08 | 5.77 | 27.77 |
| Shuffled | Counts | Observed | 184 | 52 | 51 | 38 | 19 | 36 | 11 | 391 |
| | | Expected | 230.99 | 32.58 | 32.87 | 22.69 | 41.31 | 27.06 | 3.49 | |
| | | Discrepancy | -46.99 | 19.42 | 18.13 | 15.31 | -22.31 | 8.94 | 7.51 | |
| | | Residual | -3.09 | 3.40 | 3.16 | 3.21 | -3.47 | 1.72 | 4.02 | |
| | | \|Yates Residual\| | 3.06 | 3.31 | 3.07 | 3.11 | 3.39 | 1.62 | 3.75 | |
| | | Squares | 9.36 | 10.98 | 9.45 | 9.66 | 11.52 | 2.64 | 14.07 | 67.68 |
| | | | | | | | | | Chi-square | 95.44 |
| | | | | | | | | | DF | 6 |

## Appendix 8 – Score statistics for raw and cleaned data collected on WebAssign

### Table 2-3 – Score statistics

| Score summary, data collected on WebAssign | | | | | | |
|---|---|---|---|---|---|---|
| Data type | Min. | 1st | Median | Mean | 3rd | Max. |
| raw | 0 | 7 | 10 | 10.62 | 13 | 29 |
| clean | 1 | 8 | 11 | 11.79 | 15 | 29 |
| raw PRE | 0 | 7 | 9 | 9.037 | 11 | 29 |
| clean PRE | 1 | 7 | 9 | 9.596 | 12 | 29 |
| raw POST | 0 | 8 | 12 | 13.12 | 17 | 29 |
| clean POST | 2 | 10 | 14 | 14.34 | 18.25 | 29 |

### Table 2-4 – Proportions distribution, on score quartiles

a)

| The proportion of responses in each score quartile, when all submitted responses are considered | | | |
|---|---|---|---|
| Raw data collected on WebAssign | | | |
| Score Quartile | pre-instruction (n=3703) | post-instruction (n=2374) | Pooled (n=6077) |
| 1 | 0.36 | 0.19 | 0.29 |
| 2 | 0.32 | 0.18 | 0.27 |
| 3 | 0.23 | 0.23 | 0.23 |
| 4 | 0.09 | 0.41 | 0.21 |
| Overall | 1 | 1 | 1 |

b)

| The proportion of submitted responses in each score quartile, when filters for time and multiplicity are applied | | | |
|---|---|---|---|
| Cleaned data collected on WebAssign, only test-repeaters | | | |
| Score Quartile | PRErepeat (n=1431) | POSTrepeat (n=1228) | Pooled (n=2659) |
| 1 | 0.31 | 0.12 | 0.22 |
| 2 | 0.35 | 0.18 | 0.27 |
| 3 | 0.24 | 0.24 | 0.24 |
| 4 | 0.10 | 0.46 | 0.27 |
| Overall | 1 | 1 | 1 |

c)

| Proportion differences due to response filtering (data cleaning) | | | |
|---|---|---|---|
| Data collected on WebAssign | | | |
| Score Quartile | pre-instruction ($\Delta$n=2272) | post-instruction ($\Delta$n=1146) | Pooled ($\Delta$n=3418) |
| 1 | -4.55% | -6.85% | -6.82% |
| 2 | 2.17% | 0.22% | 0.24% |
| 3 | 1.28% | 1.24% | 1.23% |
| 4 | 1.10% | 5.39% | 5.35% |

**Appendix 9 - Option choice frequency for question Q3 with fixed options.** Responses submitted for two test presentations. Cleaned data from pre-instruction (subpopulation PRE).

| Question presentation | Item-score | Option | | | | | | | | | Sums |
|---|---|---|---|---|---|---|---|---|---|---|---|
| | | A | B | C | D | E | F | G | H | I | |
| in trio (Q1,Q2,Q3) | 0 | 202 | 121 | 157 | 14 | 7 | 64 | 13 | 17 | 35 | 630 |
| | 1 | 0 | 1036 | 8 | 49 | 11 | 258 | 82 | 0 | 0 | 1444 |
| independent item | 0 | 95 | 0 | 49 | 54 | 32 | 79 | 36 | 7 | 8 | 360 |
| | 1 | 0 | 666 | 0 | 0 | 0 | 0 | 0 | 0 | 0 | 666 |